\documentclass[aps,floatfix,prd,nofootinbib,superscriptaddress,reprint,showpacs,preprintnumbers,longbibliography]{revtex4-2}
\usepackage{graphicx}
\usepackage[T1]{fontenc}
\usepackage[utf8]{inputenc}
\usepackage[UKenglish]{babel}
\usepackage{placeins}
\usepackage[binary-units]{siunitx}
\usepackage{color}
\usepackage[dvipsnames]{xcolor}
\usepackage{amsmath}
\usepackage{amssymb}
\usepackage{mathtools}
\usepackage{bbold}
\usepackage{braket}
\usepackage{feynmp-auto}
\usepackage[pdftex,
pdftitle={Real Time Simulations of Quantum Spin Chains: Density-of-States and Reweighting approaches},
pdfauthor={P. Buividovich, J. Ostmeyer},
bookmarks]{hyperref}
\usepackage{cleveref}

\graphicspath{{figures/},{plots/},{inkscape/}}

\DeclareMathOperator{\im}{i}
\DeclareMathOperator{\real}{Re}

\DeclareMathOperator{\tr}{Tr}
\DeclareMathOperator{\sinc}{sinc}

\DeclareMathOperator{\acosh}{acosh}
\DeclareMathOperator{\ord}{\mathcal{O}}

\newcommand{\lr}[1]{\left( #1 \right)}
\newcommand{\lrs}[1]{\left[ #1 \right]}
\newcommand{\expa}[1]{\exp\left( #1 \right)}

\newcommand{\matr}[2]{\left(\begin{array}{#1}#2\end{array}\right)}

\newcommand{\eto}[1]{\ensuremath{\mathrm{e}^{#1}}}
\newcommand{\md}{\ensuremath{\mathrm{d}}}

\newcommand{\id}{\ensuremath{\mathbb{1}}}
\newcommand{\ordnung}[1]{\ensuremath{\ord\left(#1\right)}}
\newcommand{\erwartung}[1]{\ensuremath{\left\langle#1\right\rangle}}

\newcommand{\mJ}{\ensuremath{\mathbb{J}}}
\newcommand{\mh}{\ensuremath{\mathbb{h}}}

\newcommand{\liverpool}{Department of Mathematical Sciences,
		University of Liverpool, United Kingdom
}

\makeatletter%
\begin{document}
\sloppy
	
	\title{Real Time Simulations of Quantum Spin Chains:\\ Density-of-States and Reweighting approaches}
	
	\author{Pavel Buividovich}
	\affiliation{\liverpool}
	\author{Johann Ostmeyer}
	\affiliation{\liverpool}
	\date{\today}
	
	\begin{abstract}
      We put the Density-of-States (DoS) approach to Monte-Carlo (MC) simulations under a stress test by applying it to a physical problem with the worst possible sign problem: the real time evolution of a non-integrable quantum spin chain. Benchmarks against numerical exact diagonalisation and stochastic reweighting are presented. Both MC methods, the DoS approach and reweighting, allow for simulations of spin chains as long as $L=40$, far beyond exact diagonalisability, though only for short evolution times $t\lesssim 1$. We identify discontinuities of the density of states as one of the key problems in the MC simulations and propose to calculate some of the dominant contributions analytically, increasing the precision of our simulations by several orders of magnitude. Even after these improvements the density of states is found highly non-smooth and therefore the DoS approach cannot outperform reweighting. We prove this implication theoretically and provide numerical evidence, concluding that the DoS approach is not well suited for quantum real time simulations with discrete degrees of freedom.
	\end{abstract}

	\maketitle

	\allowdisplaybreaks[1]
	\unitlength = 1em

\section{Introduction}

The simulation of a real-time evolution of a many-body quantum system up to some finite physical time $t$ is a problem of utmost importance across many areas of physics. However, it is an NP-hard problem which, in general, can only be solved on a classical computer in a time
\begin{eqnarray}
\label{SimTimeScaling}
 T_\text{sim} \sim \eto{\alpha\lr{t} \, n_\text{dof}}
\end{eqnarray}
that grows exponentially with the number of degrees of freedom $n_{dof}$ in a system. Polynomial-time solutions are only possible on a quantum computer \cite{Feynman:QuantMechComput, Nagaj:0908.4219}. With the current state of technology, only relatively small quantum systems can be reliably simulated on quantum computers. Therefore, at least for some time we will have to rely on classical computers to simulate the real-time evolution of quantum systems. While we certainly cannot simulate the real-time evolution in polynomial time on a classical computer, we can try to reduce the coefficient $\alpha\lr{t}$ in the exponent in (\ref{SimTimeScaling}).

Conceptually, the simplest approach to simulate real-time evolutions is to find all eigenstates of the Hamiltonian using numerical exact diagonalization. This approach works well for systems with finite-dimensional Hilbert spaces, such as quantum spin chains, and allows to calculate any real-time evolution\footnote{Up to floating-point round-off errors, that might become important at very late times.} in a straightforward way. For a chain of $L$ spin-$1/2$ degrees of freedom, such as Ising spins, the Hilbert space dimension is $N = 2^L$. Correspondingly, the computational cost of the ``plain vanilla'' exact diagonalization (ED) scales as $\ordnung{N^3} = \ordnung{2^{3L}}$. Physical results found in most of the literature scale up to $L\le18$, see e.g.~\cite{PhysRevLett.113.107204,PhysRevX.5.041047,ABANIN2021168415,PhysRevE.104.054105}. More advanced methods, such as the shift-invert method of exact diagonalization (SIMED)~\cite{10.21468/SciPostPhys.5.5.045} and polynomially filtered exact diagonalization (POLFED)~\cite{PhysRevLett.125.156601} allow for extremely efficient ED. In particular the POLFED has only $\ordnung{N^2} = \ordnung{2^{2L}}$ runtime and $\ordnung{N} = \ordnung{2^L}$ memory requirement, both up to polynomial corrections in $L$ and for a fixed number of eigenvalues. The largest systems simulated to date (to our knowledge) have $L=26$ and the largest ones used for physics beyond mere algorithmic proof of principle have $L=24$. Unfortunately the polynomial corrections are not specified in the literature but they appear to be quite significant if these are the largest sizes possible. In addition, to our knowledge it is only claimed that the limited number of eigenvalues from the bulk yield the correct results. The truncation in eigenspectrum might well cause relevant artifacts.

Another possible approach is to approximate the quantum evolution $\expa{i \hat{H} t}$ as a product of many exactly treatable factors using a Suzuki-Trotter decomposition. Instead of exact diagonalization, in this case we have to perform multiple matrix-matrix multiplications, with a computational cost that scales as $\ordnung{N^2} = \ordnung{2^{2 L}}$ for local Hamiltonians. This allows to simulate spin chains with $L\le24$ \cite{PhysRevB.103.024203,KIEFEREMMANOUILIDIS2021168481}.

An alternative to methods based on calculations in the entire Hilbert space is provided by Monte-Carlo methods, which replace exact summation over all states with importance sampling. For finite-temperature equilibrium partition functions, Monte-Carlo methods typically reduce the exponential scaling in the system size $L$ down to a polynomial one. Monte-Carlo methods can also be used for bosonic systems in a straightforward way, in contrast to exact diagonalization which necessarily relies on truncations of an infinite-dimensional Hilbert space.

In the absolute majority of cases, real-time evolution problems have path integral representations with oscillatory integrands, either real- or complex-valued. The most straightforward approach to deal with non-positive-definite integrands in Monte-Carlo simulations is reweighting, whereby one performs importance sampling with the weight proportional to the absolute value of the integrand, and re-weights the contribution of any field configuration with the corresponding complex phase of the integrand. However, in this case the NP-hardness of the real time evolution problem reveals itself in the exponentially quick deterioration of statistical power with the number of degrees of freedom $n_\text{dof}$ and the evolution time $t$. Many contributions with different complex phases cancel each other, and exponentially many Monte-Carlo samples are needed to calculate the path integral with desired precision. This exponential growth of computational complexity is referred to as the ``sign problem''.

In the past decade, the Density of States (DoS) approach has been advocated as an efficient way to make the sign problem milder \cite{Garron:1703.04649,Garron:1611.01378,Rago:1601.02929,Lucini:1910.11026}. Starting from a generic oscillatory path integral representation of a partition function $Z$ of the form
\begin{eqnarray}
\label{path_int1}
 Z = \int \mathcal{D}\phi \expa{-S_R\lrs{\phi} + \im S_I\lrs{\phi}}\,,
\end{eqnarray}
the basic idea of the DoS approach is to construct a numerical approximation to the Density of States, or probability distribution, of the complex phase $S_I\lrs{\phi}$:
\begin{eqnarray}
\label{path_int2}
 \rho\lr{E} = \int \mathcal{D}\phi \expa{-S_R\lrs{\phi}} \delta\lr{E - S_I\lrs{\phi}} \,.
\end{eqnarray}
Here $\int \mathcal{D}\phi $ denotes integration or summation over any continuous or discrete variables $\phi$. Knowing $\rho\lr{E}$, which can be determined without encountering a sign problem, we can calculate the full partition function $Z$ as a simple one-dimensional integral over $E$
\begin{eqnarray}
\label{path_int3}
 Z = \int \md E \rho\lr{E} \eto{\im E} \,.
\end{eqnarray}
An important development was the LLR method \cite{PhysRevLett.86.2050,Wang_Landau_errors,PhysRevE.78.067701} based on Robbins-Monro iterations~\cite{RobbinsMonro}, which allowed to obtain high-precision results for the Density of States $\rho\lr{E}$ for lattice models with continuous \cite{Langfeld:1509.08391,PhysRevD.102.054502} as well as discrete \cite{Guagnelli:1209.4443} degrees of freedom. Applications include the compact $U\lr{1}$ lattice gauge theory \cite{Langfeld:1509.08391}, heavy-dense QCD at finite quark density \cite{Garron:1703.04649,Garron:1611.01378}, finite-density Bose gas \cite{Rago:1601.02929}, finite-density Hubbard model \cite{PhysRevD.102.054502} and the Potts model \cite{Guagnelli:1209.4443}. It is fair to say that in all cases the computational complexity of evaluating the original multi-dimensional oscillatory path integral (\ref{path_int1}) translates into the complexity of evaluating the one-dimensional oscillatory integral (\ref{path_int3}), which is very sensitive to statistical errors in the density of states $\rho\lr{E}$. This problem was tackled in \cite{Garron:1703.04649,Garron:1611.01378,Rago:1601.02929,Guagnelli:1209.4443,PhysRevD.102.054502} by constructing various analytic approximations to $\rho\lr{E}$ and using them to calculate the integral. The approximations included polynomial fits and splines. Resummations based on Fourier transforms of $\rho\lr{E}$ were considered in \cite{PhysRevD.102.054502}. This allowed to obtain results of practical significance for lattice sizes that are intractable with other methods, e.g.\ the straightforward reweighting.

In this work, we put the DoS/LLR approach under a stress test by using it to simulate the real-time evolution, for which the partition function in (\ref{path_int1}) is strongly dominated by the imaginary part of the action $S_I$. This is hence a physical problem for which the sign problem is expected to be maximally strong. As a model of choice, we use a non-integrable quantum Ising chain with quenched disorder \cite{PhysRevLett.113.107204}. We explicitly quantify the computational cost required to obtain a fixed error of our measurements, and compare it with that of the reweighting and exact diagonalization methods. We find that without further analytic approximations, the computational cost of evaluating the one-dimensional integral (\ref{path_int3}) of the Density of States is not better than that of simple reweighting. We also critically examine the smoothness of the DoS, and find an efficient way to remove some of the dominant discontinuities, which improves the performance of both the DoS and the reweighting approaches by orders of magnitude. This also allows us to construct reasonably good polynomial approximations for the DoS $\rho\lr{E}$, however they turn out to hardly improve the quality of the results because of the high number of parameters required. Therefore, for our particular system, the DoS approach only outperforms reweighting in very rare cases. Our overall conclusion is that both Monte-Carlo approaches, LLR and reweighting, are advantageous compared to exact diagonalization when simulating short-time evolutions for large spatial system sizes, even allowing for sizes completely inaccessible to ED. On the other hand, exact diagonalization is clearly better for long times.

We chose the non-integrable quantum Ising chain with quenched disorder \cite{PhysRevLett.113.107204} as our test model because it is one of the simplest non-integrable models which exhibit ergodic and many-body localization regimes. The Hamiltonian of the model reads, in its most general form
\begin{eqnarray}
\label{Hamiltonian}
 H &= -\sigma^z \cdot \mJ \cdot \sigma^z - \mh \cdot \sigma^x ,
\end{eqnarray}
where $\sigma^{x,z}=(\sigma^{x,z}_1,\dots,\sigma^{x,z}_L)$ is the spatial vector of Pauli $x$- and $z$-spin matrices, respectively, $\mJ \in \mathbb{R}^{L\times L}$ is some coupling matrix and $\mh \in \mathbb{R}^L$ is a vector of local magnetic fields $\mh_i$.

We work in the ergodic regime, where no symmetry can accidentally make the sign problem milder. Furthermore, the model has a straightforward path integral representation. Its energy spectrum can be easily obtained using exact numerical diagonalization for chain lengths of order $L \lesssim 20$, which allows to obtain benchmark results for any real-time observables without any statistical errors. In this work we focus on one of the simplest real-time quantities, the infinite-temperature spectral form-factor
\begin{eqnarray}
\label{spf_def}
 \mathcal{K}(t) = \erwartung{K(t)}_{\mJ} =  \erwartung{\left|\tr \eto{\im H t} \right|^2}_{\mJ} ,
\end{eqnarray}
where $\erwartung{\cdot}_{\mJ}$ denotes the average over different disorder realisations and $U(t)$ is the real time evolution operator for given disorder $\mJ$. Universal late-time features of spectral form-factors, such as the ``ramp'' \cite{Yoshida:1706.05400,Hanada:1803.08050,Liu:1806.05316}, have proven useful as probes of quantum chaos in strongly-interacting many-body systems. In particular, spectral form-factors are useful for distinguishing ergodic and many-body localized regimes \cite{Prakash:2008.07547}.

Originally, our hope was to apply the DoS/LLR approach to resolve the ongoing debate about the existence and the properties of the many-body localization phase in quantum spin chains \cite{PhysRevLett.113.107204,PhysRevX.5.031032,PhysRevB.91.081103,PhysRevX.5.041047,RevModPhys.91.021001,ABANIN2021168415,PhysRevLett.95.206603,BASKO20061126,PhysRevB.95.155129,Sierant_2022,PhysRevE.104.054105,PhysRevB.103.024203,KIEFEREMMANOUILIDIS2021168481}.
The debate is strongly focused on extrapolations towards the thermodynamic limit, therefore it is useful to devise methods that could work for lattice sizes that are currently inaccessible for numerical exact diagonalization. While our improvements to the DoS/LLR approach allow to simulate spin chains of lengths as large as $L = 40$, well beyond the reach of any exact diagonalization methods, such simulations are bound to early times $t \lesssim 1$. This is by far insufficient to resolve the late-time behavior of spectral form-factors, which starts distinguishing the ergodic and the many-body localized regimes at the Heisenberg times scale $t \gtrsim 2^L$. Let us note in passing that the necessity to simulate up to times of order of $2^L$ is very likely to make the calculation of the spectral form-factor an NP-hard problem even for a quantum computer. Indeed, the ``no fast-forwarding theorem'' \cite{Atia:1610.09619} is likely to hold for our Hamiltonian. Therefore, quantum computers will need an exponentially large computational time $T_\text{sim} \gtrsim 2^L$ to simulate physical evolution up to physical times $t \gtrsim 2^L$.

The rest of this work is structured as follows. We provide the details of the model~\eqref{Hamiltonian} and its classical \mbox{$1+1$-dimensional} counterpart in \Cref{sec:model}. Next, we introduce the DoS/LLR algorithm in \Cref{sec:algorithm} and we derive how the algorithm can be optimised in \Cref{sec:approx_formulae}. Analytic estimates for the runtimes of all the different approaches are discussed in \Cref{sec:runtimes} and, finally, the results of numerical experiments are presented in \Cref{sec:results}.

\section{The model}\label{sec:model}
	
	In this work, we follow \cite{PhysRevLett.113.107204} and consider the Hamiltonian (\ref{Hamiltonian}) with both the nearest-neighbor and next-to-nearest-neighbor couplings:
\begin{eqnarray}
\label{couplings_def}
 \mJ_{ij} = \lr{\mJ_0 + \Delta \mJ_i} \, \delta_{i,j+1} + \mJ_2\delta_{i,j+2} ,
\end{eqnarray}
where the nearest neighbour coupling contains quenched disorder $\Delta \mJ_i$ that is sampled uniformly from $\Delta \mJ_i \in \left[-\Delta \mJ, \Delta \mJ\right]$. The magnetic field term $\mh_i = \mh_0$ is constant. Throughout this work we choose $\mh_0=\num{0.6}$, $\mJ_2=\num{0.3}$, $\mJ_0=\num{1}$, and $\Delta \mJ=\num{1}$ deeply in the chaotic regime.
	
The real time dynamics of the system as well as the severity of its sign problem are captured in the spectral form factor (SFF)
\begin{align}
	\mathcal{K}(t) &\equiv \erwartung{K(t)}_{\mJ}\\
	&\equiv \erwartung{\left|\tr U(t)\right|^2}_{\mJ}
\end{align}
	where $\erwartung{\cdot}_{\mJ}$ denotes the average over different disorder realisations and $U(t)$ is the real time evolution operator for given disorder $\mJ$
	\begin{align}
		\begin{split}
			U(t) &= \eto{-\im Ht}
			=
			\prod_{j=1}^{N_t} \eto{-\im \delta H}
			\\
			&= \prod_{j=1}^{N_t} \eto{\im \delta \sigma^z \cdot \mJ\cdot \sigma^z}\, \eto{\im \delta \mh\cdot \sigma^x} + \ordnung{\delta}
		\end{split}\label{eq:trotter_decomposition}
	\end{align}
	with the physical evolution time $t$ and the Suzuki-Trotter discretization step size $\delta \equiv t/N_t$. This Suzuki-Trotter decomposition becomes exact in the limit where the number of time steps $N_t \rightarrow \infty$.
	
	The transfer-matrix method allows to relate the Trotterised version to a two-dimensional system of classical spins\footnote{Such a system can also be interpreted as an anisotropic classical Ising model with complex coupling coefficients.}
	\begin{align}
		K(t) &\propto Z \equiv \sum_{\{s\}} \eto{-S(s)}\,, \label{eq:partition_sum_all_spins}
	\end{align}
	where we sum over all possible spin $s_{i,k}=\pm1$ combinations, arriving at the action
    \begin{align}
    	\begin{split}
        S &= -\im\sum_k\sum_{i,j}s_{i,k}J_{ij}s_{j,k}
		\\ 
        &\quad -\sum_i\sum_k\left(h_i - \frac\pi4\im\right) s_{i,k}s_{i,k+1} \, ,
        \end{split}\\
	J_{ij} &\coloneqq \delta \mJ_{ij} \, , \quad  h_i \coloneqq -\frac12 \log\tan\lr{\delta \mh_i} \, .
    \end{align}
	$J$ and $h_i$ are real constants. See \cref{sec:quantum_to_classical} for details on the derivation.
	
	Any spin flip results in a change of $S$ by $\lambda \left(h_i - \frac\pi4\im\right)$ with $\lambda \in \{\pm4, 0\}$ and some contributions of order $\delta$. Therefore both, real and imaginary part, are significantly changed even for small $\delta$. In particular, $\lambda=\pm4$ flips the sign of the action.  	%

	We split the action into three separate parts
	\begin{align}
		S &= \im S_I + S_R + \im \pi S_S\,,\label{eq:split_action}\\
		S_I &\coloneqq -\sum_k\sum_{i,j}s_{i,k} J_{ij}s_{j,k}\\
		S_R &\coloneqq - \sum_i\sum_k h_i s_{i,k}s_{i,k+1}\\
		S_S &\coloneqq \frac14 \sum_i\sum_k s_{i,k}s_{i,k+1}\,.
	\end{align}
	Now each of $S_I$, $S_R$ and $S_S$ is real. Furthermore $S_S$ is an integer and therefore
	\begin{align}
		\eto{\im\pi S_S} &= \pm 1
	\end{align}
	contributes only a sign to the total measure.
	
	\subsection{Probability weighted density of states}
	Monte-Carlo based algorithms can easily implement the probability density induced by $S_R$ whereas the complex phase and real sign contributions by $S_I$ and $S_S$ respectively pose severe issues. As outlined in equations (\ref{path_int1})-(\ref{path_int3}) in the introductory Section, in this work we employ the density of states (DoS) approach, rewriting the partition sum as
	\begin{align}
		Z &= \sum_{\{s\}} \eto{-S_R} \eto{-\im S_I} \eto{-\im \pi S_S}\\
		&= \sum_{\{s\}} \eto{-S_R} \int_{\mathbb{R}} \md E\, \eto{-\im E}\,\delta(E - S_I) \sum_{z=\pm1}z\,\delta_{z,\eto{-\im \pi S_S}}\\
		&= \sum_{z=\pm1} z \int_{\mathbb{R}} \md E\, \eto{-\im E} \sum_{\{s\}} \eto{-S_R}\, \delta(E - S_I)\, \delta_{z,\eto{-\im \pi S_S}}\\
		&= \sum_{z=\pm1} z \int_{\mathbb{R}} \md E\, \eto{-\im E}\, \rho_z(E)\,.\label{eq:partition_sum_dos}
	\end{align}
	In the last step we defined the DoS $\rho_z\lr{E}$ as
	\begin{align}
		\rho_z(E) \coloneqq \sum_{\{s\}} \eto{-S_R}\, \delta(E - S_I)\, \delta_{z,\eto{-\im \pi S_S}}\,.\label{eq:dos_definition}
	\end{align}

	\subsection{Normalisation}
	Let us define the auxiliary partition sum
	\begin{align}
		Z_R &= \sum_{\{s\}} \eto{-S_R}\\
		&= \sum_{z=\pm1} \int_{\mathbb{R}} \md E\, \rho_z(E)
	\end{align}
	which is a sum of manifestly positive contributions and therefore does not suffer from a sign problem. Note that all terms containing the coupling $\mJ$ have been dropped and $Z_R$ therefore completely decouples into its one-dimensional (time-like) parts, i.e.
	\begin{align}
		Z_R &= \prod_i Z_{R,i}\,,\\
		Z_{R,i} &= \tr \left[(T_2)_i^{N_t}\right]\label{eq:trace_partition_sum}\\
		&=2^{N_t}\left(\cosh^{N_t}  h_i + \sinh^{N_t}  h_i\right)\,,
	\end{align}
	where the last line is the well known solution of the one-dimensional classical Ising model and can be obtained by diagonalising the $2 \times 2$ transfer matrices $(T_2)_i$ from equation~\eqref{eq:temporal_transfer_matrix}.
	
	In the full simulation one would usually normalise the sum over the DoS to one so that the Monte-Carlo estimator (denoted by $\erwartung{\cdot}$) of the partition sum becomes
	\begin{align}
		\erwartung{Z} &= Z_R \erwartung{\frac{Z}{Z_R}}\,,
	\end{align}
	where $Z_R$ is known analytically and $\erwartung{\frac{Z}{Z_R}}$ can be found by evaluating equation~\eqref{eq:partition_sum_dos} with the normalised estimator of $\rho_z(E)$. Combining this result with equation~\eqref{eq:classical_to_quantum}, we finally obtain a formula for the trace of the quantum mechanical time evolution operator
	\begin{align}
		\tr U(t) &= A Z_R \erwartung{\frac{Z}{Z_R}}\,.
	\end{align}  	%
\section{The algorithms}\label{sec:algorithm}
	The canonical choice and baseline algorithm for Monte-Carlo simulations with sign problem is \textit{reweighting}. In this approach all contributions stemming from the imaginary part of the action are simply considered part of the observable. In this particular case this means that configurations are sampled purely from the probability distribution induced by $S_R$ (using for instance local Metropolis-Hastings updates) and the partition sum is estimated by
	\begin{align}
		\erwartung{\frac{Z}{Z_R}} &= \erwartung{\eto{-\im S_I-\im\pi S_S}}_{S_R}\,.
	\end{align}
	
	Another natural approach is to estimate the DoS $\rho$ prior to determining any observables. This can be done efficiently using the \textit{Logarithmic Linear Relaxation} (LLR) algorithm.
	The principle idea of LLR was first proposed by Wang and Landau~\cite{PhysRevLett.86.2050} and applied to the classical Ising model with uniform coupling in two dimensions. Later on it has been optimised through the introduction of the $1/t$ algorithm~\cite{Wang_Landau_errors,PhysRevE.78.067701} in accordance with the theoretically optimal scheme derived by Robbins and Monro~\cite{RobbinsMonro}.
	
	Let us consider a classical Ising system of $N\equiv LN_t$ spins. The auxiliary variable $E$  (typically identified as an `energy', see eq.~\eqref{eq:dos_definition}) is divided into $M$ bins $E_i$, $i=1,\dots,M$, so that we discretise the DoS
	\begin{align}
		\rho_i \coloneqq \int_{\frac{E_i+E_{i-1}}{2}}^{\frac{E_i+E_{i+1}}{2}}\md E \,\rho(E)\,,\quad i=2,\dots,M-1\label{eq:dos_to_bins}
	\end{align}
	and for $i=1$ ($i=M$) the lower (upper) bound of the integral takes the minimal (maximal) value of $E$. Of course, the arithmetic mean can be exchanged for any other value between $E_i$ and its neighbours without changing the general argument. Discretisation errors are of order $\ordnung{h^2}$ and further details are discussed in \cref{sec:error_with_bins}.
	
	We write
	\begin{align}
		\rho_i\equiv\eto{\alpha_i}
	\end{align}
	and from now on we aim to estimate the $\alpha_i$ as accurately as possible (hence the `logarithmic' in the method's name). Starting from uniform initial conditions ($\alpha_i=0$ for all $i$), we employ an update scheme of $\Lambda$ steps in total where at the $k$th step some configuration update $s\mapsto s'$ is proposed and accepted with the probability
	\begin{align}
		p_\text{acc} &= \frac{\eto{-S_R\lrs{s'}}}{\eto{-S_R\lrs{s}}}\cdot\frac{\rho_{i(s)}}{\rho_{i(s')}}\label{eq:acc_prob}\\
		&= \eto{S_R\lrs{s}-S_R\lrs{s'}}\,\eto{\alpha_{i(s)}-\alpha_{i(s')}}\,,
	\end{align}
	where $i\lr{s}$ and $i\lr{s'}$ are the bins that correspond to the imaginary part of the action $S_I$ on configurations $s$ and $s'$, respectively.
	The DoS of the accepted bin, say $i$ w.l.o.g., has to be updated
	\begin{align}
		\alpha_i \mapsto \alpha_i + \beta_{N,M,\Lambda}(k,i)\,,
        \label{eq:robins_monro}
	\end{align}
	where $\beta \ge 0$. Without prior knowledge of the DoS we drop the explicit dependence on $i$. Furthermore we know from Ref.~\cite{RobbinsMonro} that the conditions
	\begin{align}
		\sum_{k=0}^\infty \beta_{N,M,\Lambda}(k) &= \infty\,,\label{eq:beta_sum_infinite}\\
		\sum_{k=0}^\infty \beta_{N,M,\Lambda}(k)^2 &< \infty
	\end{align}
	have to hold.\footnote{Actually these are not exactly the conditions from~\cite{RobbinsMonro}, but they are equivalent to this simpler version.} In particular, they propose a function asymptotically scaling as $1/k$ since this is the fastest decaying function (up to logarithmic factors) of the required class and therefore promises best convergence for large $k$.
	
	The coefficients in the explicit form
	\begin{align}
		\beta_{N,M,\Lambda}(k) &= \frac{a_{N,M,\Lambda}}{b_{N,M,\Lambda}+k}
	\end{align}
	are of crucial importance and we derive in \cref{sec:llr_params} why
	\begin{align}
		a_{N,M,\Lambda} &= \ln2\,\frac{MN}{\ln\frac{\Lambda}{b_{N,M,\Lambda}}}\,,\label{eq:opt_a}\\
		b_{N,M,\Lambda} &= 3M\label{eq:opt_b}
	\end{align}
	are a particularly good choice. We are going to use them throughout the rest of this work.
	
	LLR iterations do not form a Markov chain because the accept/reject probability depends on the history of the previous updates. Therefore statistical errors cannot be calculated via straight forward bootstrap procedures or similar. Instead every simulation presented hereafter has been repeated $N_\text{runs}=40$ times and the error has been estimated from the resulting distribution. In practice LLR simulations are not always stable but are prone to outliers. To counteract this effect we used the median as best estimator for observables instead of the more commonly used mean. Similarly the errors have been approximated using the 16\% and 84\% quantiles instead of the standard deviation.
	For a normal distribution this choice of median and quantiles is fully compatible with the default choice of mean and standard deviation, but for other distributions it is more stable.  	%
	\section{Small transverse component expansion}\label{sec:approx_formulae}

	The representation (\ref{eq:partition_sum_all_spins}) of the real-time partition function in terms of a finite sum over classical Ising spins implies a finite number of states, so that technically speaking the DoS (\ref{eq:dos_definition}) is a collection of delta peaks. However, in the presence of quenched disorder, there are in general $\ordnung{2^{L}}$ delta peaks spread randomly over the range of width $\ordnung{L}$ of possible values of $S_I$. As a result, the bulk of the DoS approaches the continuous function in the thermodynamic limit $L \rightarrow +\infty$. For finite length $L$, the apparent continuousness is a mere consequence of the finite resolution following from the fixed number $M$ of bins defining the DoS, i.e.\ the true distribution is convoluted with a function that is $1$ within a bin and $0$ everywhere else.
	
\begin{figure*}[ht]
		\centering
		\includegraphics[width=.45\textwidth]{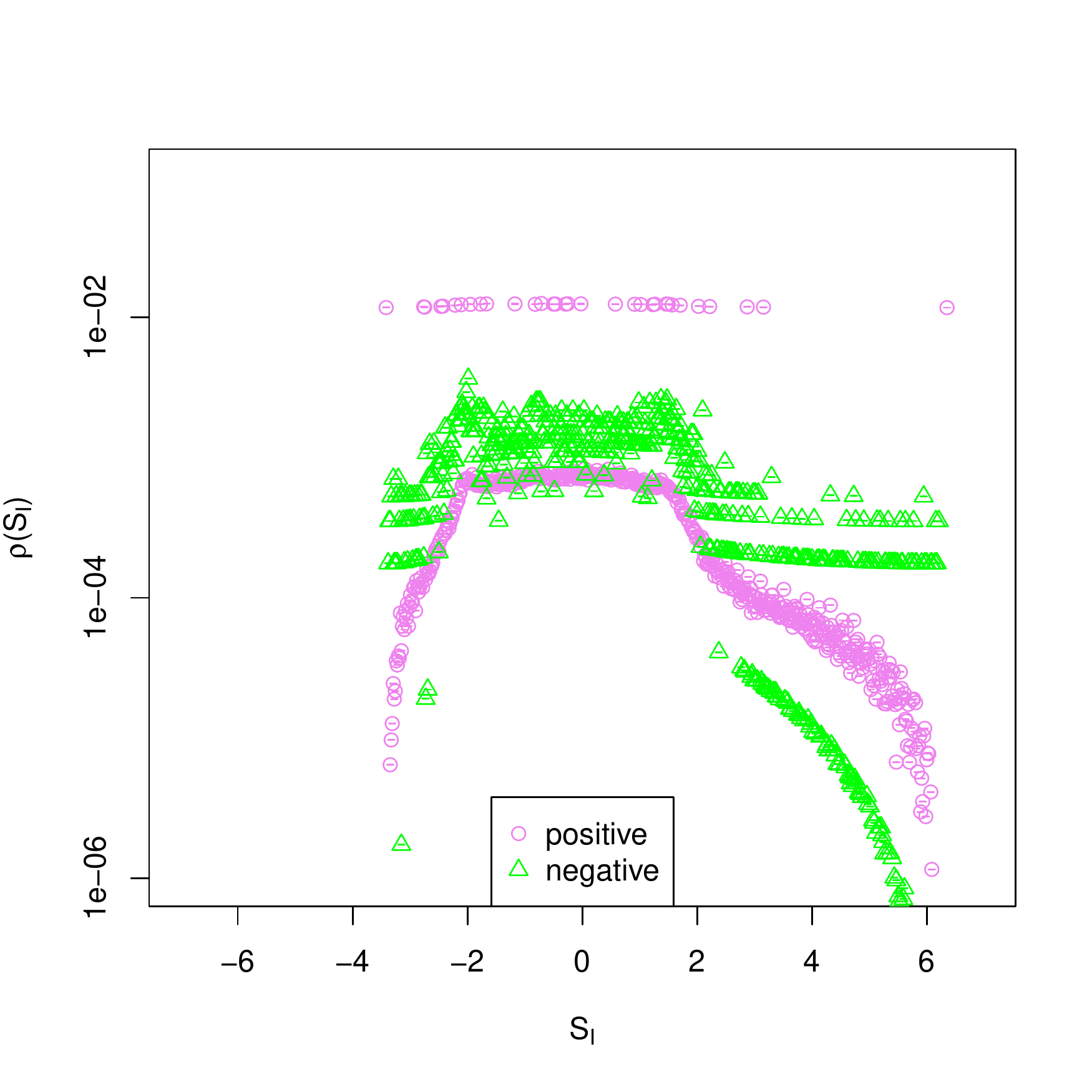}
		\includegraphics[width=.45\textwidth]{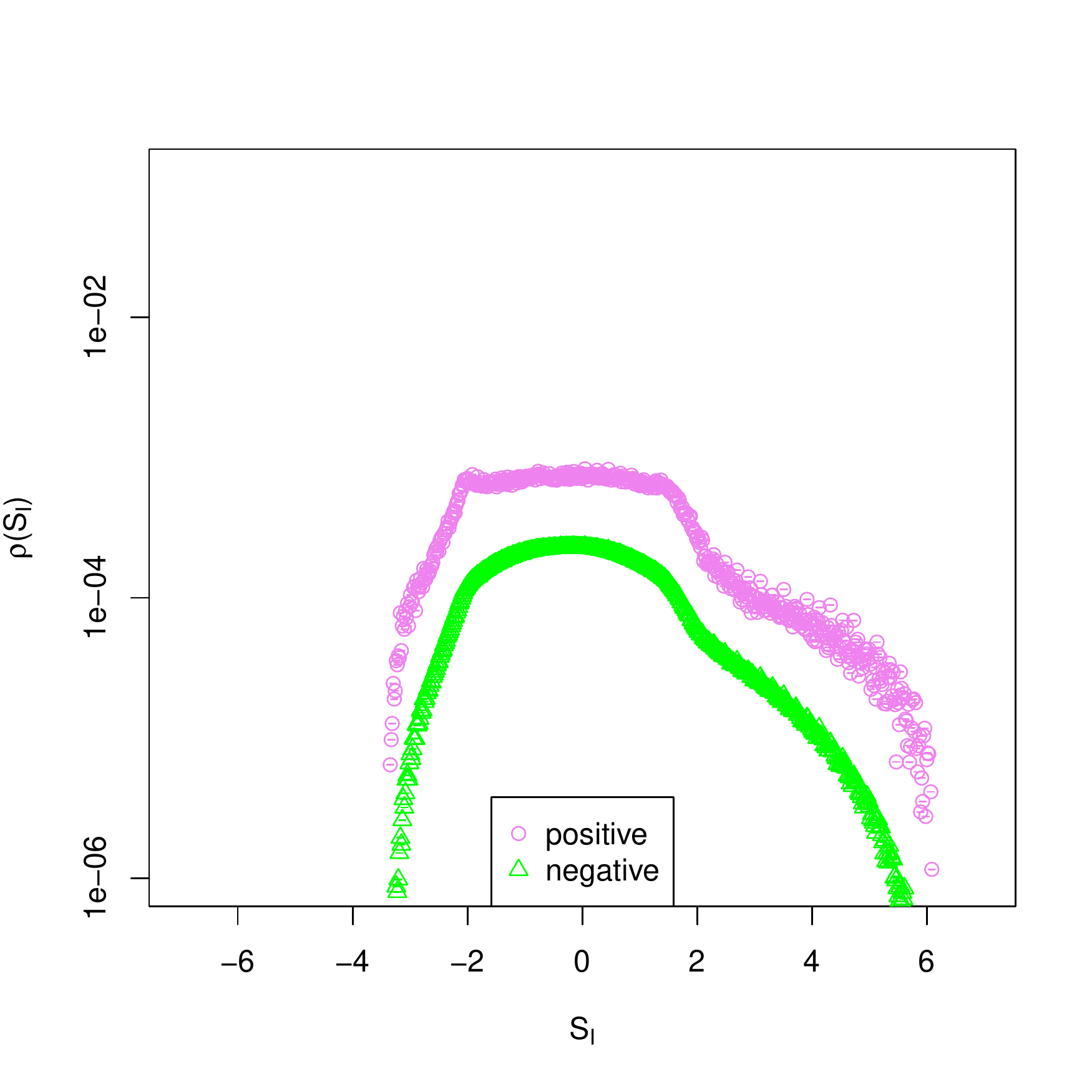}
		\caption{Densities of states in the even (positive) and odd (negative) parity sectors respectively, calculated with LLR$_0$ (left) and LLR$_2$ (right). The spin chain length is $L=6$, the evolution time is $t=1$, the number of Monte-Carlo sweeps is $N_s=10^8$.}
    \label{fig:dos_llr0_vs_llr2}
	\end{figure*}

	Looking at a typical distribution of the DoS (see fig.~\ref{fig:dos_llr0_vs_llr2}), it appears to be fractured into bands. The top one corresponds to the even sector and consists of a few very narrow peaks. The second (odd) band is closer to a smooth distribution, though it still features some irregularities. An alternating descend of smooth bands with roughly equal step size on a logarithmic scale follows. With the technically available precision we usually cannot resolve any but the four highest bands.
	
	It is straight forward to interpret these bands, as they directly correspond to a particular parity sector each. Every spin-flip pair in temporal direction (see fig.\ref{fig:flip-pairs}) changes this sector. It comes with a sign flip and a Boltzmann suppression, i.e.
	\begin{align}
		S_S & \mapsto S_S - 1\,,\\
		S_R & \mapsto S_R + 4h\,,\\
		\Rightarrow p & \mapsto -p\left(\tan \frac{t \mh}{N_t}\right)^2\,,
	\end{align}
	where $p=\eto{-S_R-\im\pi S_S}$ is the sign-full `probability' weight and $S_{S,R}$ as in equation~\eqref{eq:split_action}. This provides a canonical way of expanding the DoS $\rho_z(E)$ and the partition sum $Z$ in powers of $\tan \frac{t \mh}{N_t}$, that is around small times $t$ or transverse fields $\mh$.

	From now on we will denote simulations restricted to parity sectors with at least $n$ spin-flip pairs by LLR$_n$ (or REW$_n$ for reweighting). In practice this means that the acceptance probability in equation~\eqref{eq:acc_prob} is set to zero for any proposed configuration with less than $n$ spin-flip pairs. Otherwise the sampling algorithm proceeds unchanged.
	
	Thus, the ``plain vanilla'' algorithm allowing to visit the full phase space will be denoted LLR$_0$ and a lot of our studies will be focused on LLR$_2$ as a good trade off between precision and performance. The remaining contributions from all the configurations with less than $n$ spin-flip pairs are calculated analytically as follows.
	
	\subsection{Leading order: \texorpdfstring{$n=0$}{n=0} spin-flip pairs}
    \label{subsec:leading_order}

	Without any spin flips in temporal direction the phase space is confined to one of the $2^L$ spatial configurations repeated $N_t$ times. Therefore, the DoS and the classical partition sum reduce to
	\begin{align}
		\left.\rho_z(E)\right|_{n=0} &\propto \delta_{z,1}\, \sum_{\{s\}}\delta\left(E-t \sum_{i,j}s_i\mJ_{ij}s_j\right)\,,\\
		\left.Z\right|_{n=0} &= \sum_{\{s\}}\eto{\im N_t \sum_{i,j}s_i J_{ij}s_j + LN_t h}\\
		&= \left(\tan\frac{t \mh}{N_t}\right)^{-\frac 12 LN_t} \sum_{\{s\}} \eto{\im t \sum_{i,j}s_i\mJ_{ij}s_j}
	\end{align}
	and thus
	\begin{align}
		\left.\tr U(t)\right|_{n=0} &= \sqrt{\frac{\sin\delta \mh\cos\delta \mh}{\tan \delta \mh}}^{LN_t} \sum_{\{s\}} \eto{\im t \sum_{i,j}s_i \mJ_{ij} s_j}\\
		&= \left(\cos\delta \mh\right)^{LN_t} \sum_{\{s\}} \eto{\im t \sum_{i,j}s_i \mJ_{ij} s_j}\,.\label{eq:no_flip}
	\end{align}
	The prefactor $Z_0\equiv\left(\cos\delta \mh\right)^{LN_t} \overset{\delta\rightarrow0}{\longrightarrow}1$ is clearly an artefact from Trotterisation, but it cannot be dropped in realistic simulation since it crucially counteracts finite $\delta$ effects in the LLR$_n$ simulation itself.

	Here we assume that the couplings $\mJ_{ij}$ are non-uniform and hence the full sum over all configurations $\{s\}$ has to be calculated. In particular simpler cases, e.g.\ $\mJ_{ij}=\mJ\,\delta_{i+1,j}$, the sum can be simplified and therefore the computational complexity reduced dramatically.
	
	Note that the continuum limit of this result corresponds to the $\mh=0$ approximation of the quantum mechanical time evolution operator
	\begin{align}
		\tr U(t) &= \tr \left[\eto{\im t \sigma^z \cdot \mJ \cdot \sigma^z}\,\eto{\sigma^x \ordnung{ht}}\right]\\
		&= \sum_{\{s\}} \eto{\im t \sum_{i,j}s_i\mJ_{ij}s_j} + \ordnung{\left(ht\right)^2}\,,
	\end{align}
	where the linear order in $ht$ can be dropped because $\tr\sigma^x=0$.
	
	\subsection{Next-to-leading odd order: \texorpdfstring{$n=1$}{n=1} spin-flip pair}
    \label{subsec:next_to_leading_order}

	\begin{figure*}[ht]
		\centering
\begingroup%
  \makeatletter%
  \providecommand\color[2][]{%
    \errmessage{(Inkscape) Color is used for the text in Inkscape, but the package 'color.sty' is not loaded}%
    \renewcommand\color[2][]{}%
  }%
  \providecommand\transparent[1]{%
    \errmessage{(Inkscape) Transparency is used (non-zero) for the text in Inkscape, but the package 'transparent.sty' is not loaded}%
    \renewcommand\transparent[1]{}%
  }%
  \providecommand\rotatebox[2]{#2}%
  \newcommand*\fsize{\dimexpr\f@size pt\relax}%
  \newcommand*\lineheight[1]{\fontsize{\fsize}{#1\fsize}\selectfont}%
  \ifx\svgwidth\undefined%
    \setlength{\unitlength}{115.02194539bp}%
    \ifx\svgscale\undefined%
      \relax%
    \else%
      \setlength{\unitlength}{\unitlength * \real{\svgscale}}%
    \fi%
  \else%
    \setlength{\unitlength}{\svgwidth}%
  \fi%
  \global\let\svgwidth\undefined%
  \global\let\svgscale\undefined%
  \makeatother%
  \begin{picture}(1,0.89667689)%
    \lineheight{1}%
    \setlength\tabcolsep{0pt}%
    \put(0,0){\includegraphics[width=\unitlength,page=1]{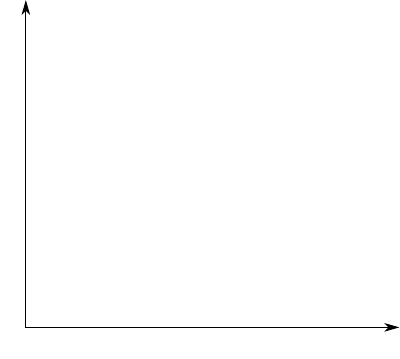}}%
    \put(0.94203539,0.00415278){\color[rgb]{0,0,0}\makebox(0,0)[lt]{\lineheight{1.25}\smash{\begin{tabular}[t]{l}$x$\end{tabular}}}}%
    \put(-0.00234543,0.84993489){\color[rgb]{0,0,0}\makebox(0,0)[lt]{\lineheight{1.25}\smash{\begin{tabular}[t]{l}$t$\end{tabular}}}}%
    \put(0,0){\includegraphics[width=\unitlength,page=2]{0-flips.pdf}}%
  \end{picture}%
\endgroup%
 		\hfill
\begingroup%
  \makeatletter%
  \providecommand\color[2][]{%
    \errmessage{(Inkscape) Color is used for the text in Inkscape, but the package 'color.sty' is not loaded}%
    \renewcommand\color[2][]{}%
  }%
  \providecommand\transparent[1]{%
    \errmessage{(Inkscape) Transparency is used (non-zero) for the text in Inkscape, but the package 'transparent.sty' is not loaded}%
    \renewcommand\transparent[1]{}%
  }%
  \providecommand\rotatebox[2]{#2}%
  \newcommand*\fsize{\dimexpr\f@size pt\relax}%
  \newcommand*\lineheight[1]{\fontsize{\fsize}{#1\fsize}\selectfont}%
  \ifx\svgwidth\undefined%
    \setlength{\unitlength}{115.02194539bp}%
    \ifx\svgscale\undefined%
      \relax%
    \else%
      \setlength{\unitlength}{\unitlength * \real{\svgscale}}%
    \fi%
  \else%
    \setlength{\unitlength}{\svgwidth}%
  \fi%
  \global\let\svgwidth\undefined%
  \global\let\svgscale\undefined%
  \makeatother%
  \begin{picture}(1,0.89667689)%
    \lineheight{1}%
    \setlength\tabcolsep{0pt}%
    \put(0,0){\includegraphics[width=\unitlength,page=1]{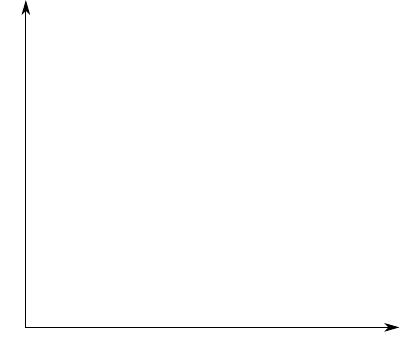}}%
    \put(0.94203539,0.00415278){\color[rgb]{0,0,0}\makebox(0,0)[lt]{\lineheight{1.25}\smash{\begin{tabular}[t]{l}$x$\end{tabular}}}}%
    \put(-0.00234543,0.84993489){\color[rgb]{0,0,0}\makebox(0,0)[lt]{\lineheight{1.25}\smash{\begin{tabular}[t]{l}$t$\end{tabular}}}}%
    \put(0,0){\includegraphics[width=\unitlength,page=2]{1-flip.pdf}}%
  \end{picture}%
\endgroup%
 		\hfill
\begingroup%
  \makeatletter%
  \providecommand\color[2][]{%
    \errmessage{(Inkscape) Color is used for the text in Inkscape, but the package 'color.sty' is not loaded}%
    \renewcommand\color[2][]{}%
  }%
  \providecommand\transparent[1]{%
    \errmessage{(Inkscape) Transparency is used (non-zero) for the text in Inkscape, but the package 'transparent.sty' is not loaded}%
    \renewcommand\transparent[1]{}%
  }%
  \providecommand\rotatebox[2]{#2}%
  \newcommand*\fsize{\dimexpr\f@size pt\relax}%
  \newcommand*\lineheight[1]{\fontsize{\fsize}{#1\fsize}\selectfont}%
  \ifx\svgwidth\undefined%
    \setlength{\unitlength}{115.02194539bp}%
    \ifx\svgscale\undefined%
      \relax%
    \else%
      \setlength{\unitlength}{\unitlength * \real{\svgscale}}%
    \fi%
  \else%
    \setlength{\unitlength}{\svgwidth}%
  \fi%
  \global\let\svgwidth\undefined%
  \global\let\svgscale\undefined%
  \makeatother%
  \begin{picture}(1,0.89667689)%
    \lineheight{1}%
    \setlength\tabcolsep{0pt}%
    \put(0,0){\includegraphics[width=\unitlength,page=1]{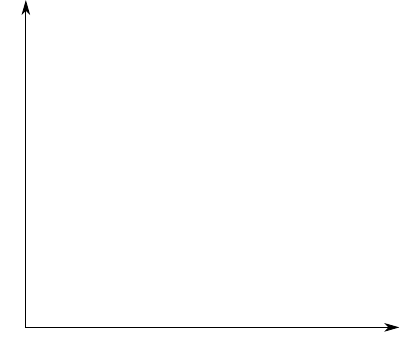}}%
    \put(0.94203539,0.00415278){\color[rgb]{0,0,0}\makebox(0,0)[lt]{\lineheight{1.25}\smash{\begin{tabular}[t]{l}$x$\end{tabular}}}}%
    \put(-0.00234543,0.84993489){\color[rgb]{0,0,0}\makebox(0,0)[lt]{\lineheight{1.25}\smash{\begin{tabular}[t]{l}$t$\end{tabular}}}}%
    \put(0,0){\includegraphics[width=\unitlength,page=2]{1p-flip.pdf}}%
  \end{picture}%
\endgroup%
 		\hfill
\begingroup%
  \makeatletter%
  \providecommand\color[2][]{%
    \errmessage{(Inkscape) Color is used for the text in Inkscape, but the package 'color.sty' is not loaded}%
    \renewcommand\color[2][]{}%
  }%
  \providecommand\transparent[1]{%
    \errmessage{(Inkscape) Transparency is used (non-zero) for the text in Inkscape, but the package 'transparent.sty' is not loaded}%
    \renewcommand\transparent[1]{}%
  }%
  \providecommand\rotatebox[2]{#2}%
  \newcommand*\fsize{\dimexpr\f@size pt\relax}%
  \newcommand*\lineheight[1]{\fontsize{\fsize}{#1\fsize}\selectfont}%
  \ifx\svgwidth\undefined%
    \setlength{\unitlength}{115.02194539bp}%
    \ifx\svgscale\undefined%
      \relax%
    \else%
      \setlength{\unitlength}{\unitlength * \real{\svgscale}}%
    \fi%
  \else%
    \setlength{\unitlength}{\svgwidth}%
  \fi%
  \global\let\svgwidth\undefined%
  \global\let\svgscale\undefined%
  \makeatother%
  \begin{picture}(1,0.89667689)%
    \lineheight{1}%
    \setlength\tabcolsep{0pt}%
    \put(0,0){\includegraphics[width=\unitlength,page=1]{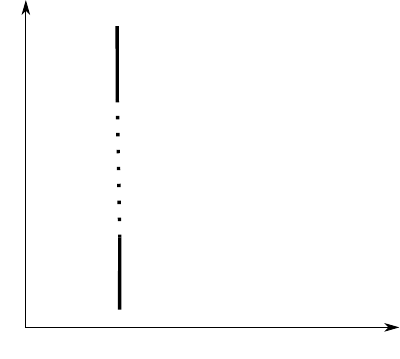}}%
    \put(0.94203539,0.00415278){\color[rgb]{0,0,0}\makebox(0,0)[lt]{\lineheight{1.25}\smash{\begin{tabular}[t]{l}$x$\end{tabular}}}}%
    \put(-0.00234543,0.84993489){\color[rgb]{0,0,0}\makebox(0,0)[lt]{\lineheight{1.25}\smash{\begin{tabular}[t]{l}$t$\end{tabular}}}}%
    \put(0,0){\includegraphics[width=\unitlength,page=2]{4-flips.pdf}}%
  \end{picture}%
\endgroup%
 		\caption{Visualisation of configurations with different numbers $n$ of spin-flip pairs in the time continuum. Periodic boundary conditions are applied, so flips come in pairs only. W.l.o.g.\ the solid lines represent spin-up and the dotted lines spin-down. From left to right: $n=0$, $n=1$, $n=1$, $n=4$.}\label{fig:flip-pairs}
	\end{figure*}

	Let us now consider the case of a single spin-flip pair, that is exactly one of the quantum spins has exactly two different values over time in the $z$-basis. Some visualisations can be found in figure~\ref{fig:flip-pairs}. We absorb the probability weight $-\left(\tan\delta \mh\right)^2$ and the combinatorial factor $N_t/2$ (temporal translational invariance yields a factor $N_t$, exchangeability of start and end point of the flipped region a factor $1/2$) into a new prefactor
	\begin{align}
		Z_1 &\equiv -\frac{N_t}{2}\left(\tan\delta \mh\right)^2 Z_0\,.
	\end{align}
	Now for every configuration $s$ without flips there are $L$ different candidates $l$ for a flip with $N_t-1$ different possible lengths $\tau$ each. We can write the resulting phase as
	\begin{eqnarray}
		\left.S_I\right|_{n=1} = -\left(N_t - \tau\right)\sum_{i,j}s_i  J_{ij} s_j
        - \nonumber \\ -
	    \tau\sum_{i,j} s_i\left(1-2\delta_{i l}\right) J_{ij} \left(1-2\delta_{ l j}\right)s_j
		= \nonumber \\ =
		-\left(t\sum_{i,j}s_i \mJ_{ij} s_j  - 4\tau\sum_j s_l  J_{ l j}s_j\right)\,.
	\end{eqnarray}

	\begin{widetext}
	Combining the pre-factor and the phase leads to
	\begin{align}
		\left.\rho_z(E)\right|_{n=1} &= Z_1\, \delta_{z,-1}\, \sum_{\{s\}} \sum_l \sum_{\tau=1}^{N_t-1} \delta\left(E-t \sum_{i,j}s_i\mJ_{ij}s_j + 4\tau\sum_j s_l  J_{ l j}s_j\right)\,,\\
		\left.\tr U(t)\right|_{n=1} &= Z_1\sum_{\{s\}}\eto{\im t \sum_{i,j}s_i\mJ_{ij}s_j} \sum_l \sum_{\tau=1}^{N_t-1}\eto{-4\im \tau \sum_j s_l  J_{l j}s_j}\label{eq:tr_U_n1}\\
		\begin{split}
			&= Z_1\sum_{\{s\}}\eto{\im t \sum_{i,j}s_i\mJ_{ij}s_j} \sum_l \eto{-2\im t \sum_j s_l \mJ_{l_j}s_j}\frac{\sin\left(2\left(N_t-1\right)\sum_j s_l J_{l j} s_j\right)}{\sin\left(2\sum_j s_l J_{l j} s_j\right)}
		\end{split}
	\end{align}
	with the well-defined continuum limit
	\begin{align}
		\left.\rho_z(E)\right|_{n=1} &= \frac{(ht)^2}{2}\, \delta_{z,-1}\, \sum_{\{s\}} \sum_l \theta\left(-\left|E - t \sum_{i,j}s_i\mJ_{ij}s_j\right| + \left|4\tau\sum_j s_l  J_{ l j}s_j\right|\right)\,,\label{eq:cont_rho_n1}\\
		\begin{split}
		\lim\limits_{\delta\rightarrow0}\left.\tr U(t)\right|_{n=1} &= -\frac{(ht)^2}{2}\sum_{\{s\}}\eto{\im t \sum_{i,j}s_i\mJ_{ij}s_j}\sum_l \eto{-2\im t \sum_j s_l \mJ_{l_j}s_j} \sinc\left(2t\sum_j s_l \mJ_{l j} s_j\right)\,,
		\end{split}\label{eq:single_flip}
	\end{align}
	where $\theta$ denotes the Heaviside step function.
	\end{widetext}

	Note that the temporal sum in~\eqref{eq:tr_U_n1} could be calculated exactly since it reduced to a geometric series. This allows an evaluation of the remaining sum in $\ordnung{L2^L}$, i.e.\ the same runtime as the contribution without flips.
	
	Again we could have obtained the continuum result up to leading order directly from the time evolution operator using the Zassenhaus formula
	\begin{align}
		\tr U(t) &= \tr\left[\eto{\im t \sigma^z \cdot \mJ \cdot \sigma^z}\eto{\im t \mh\sum_i\sigma^x_i}\right] + \ordnung{(ht)^4}\\
		\begin{split}
		&= \tr\left[\eto{\im t \sigma^z \cdot \mJ \cdot \sigma^z}\prod_i\left(\id+\im ht\sigma^x_i - \frac12(ht)^2\id\right)\right]\\
		&\qquad + \ordnung{(ht)^4}
		\end{split}\\
		&= \left(1-\frac12 L (ht)^2\right)\tr\left[\eto{\im t \sigma^z \cdot \mJ \cdot \sigma^z}\right] + \ordnung{(ht)^4}\,.\label{eq:first_order_trotter}
	\end{align}
	Here the cyclic property of the trace ensures that the simple first-order decomposition into $\sigma^z$ and $\sigma^x$ parts is equivalent to a symmetric decomposition with an error of order $\ordnung{(ht)^3}$. Additionally $\tr(\sigma^x)^3=0$ reduces the residual error to $\ordnung{(ht)^4}$.
	
	\subsection{Higher orders: \texorpdfstring{$n\ge2$}{n>=2} spin-flip pairs}
	In principle, we can follow the steps of the previous section in order to write down the contribution of the $n$th order ($n$ spin flips) for arbitrary $n$. The prefactor (without combinatorics) reads
	\begin{align}
		Z_n &\equiv \left(-\tan^2\delta \mh\right)^n
	\end{align}
	and the phase (including combinatorics) generalises to
	\begin{align}
		\begin{split}
			S_I &= \sum_k \sum_{i,j} s_i\left(\prod_l\left(1-2\delta_{il}\theta(k-\tau^l_1)\theta(\tau^l_2-k)\right)\right)\\
			&\qquad\times   J_{ij}\left(\prod_l\left(1-2\delta_{l j}\theta(k-\tau^l_1)\theta(\tau^l_2-k)\right)\right) s_j
		\end{split}
	\end{align}
	where $\theta$ denotes the Heaviside step function and $\tau^l_{1,2}$ are successive flipping times of spins at spatial position $l$. The partition sum has to include all possible combinations of $\{l\}$ and corresponding $\tau^l_{1,2}$.
	
	The evaluation of the partition sum for any $n\ge2$ however is prohibitively expensive because the temporal sum can no longer be evaluated analytically. Two different regions flipped at overlapping times introduce non-trivial entanglement (see right panel of fig.~\ref{fig:flip-pairs}). Another way to see this is the fact that the first-order Trotter decomposition of the time evolution operator results in an error of order $\ordnung{(ht)^4}$. A higher order treatment therefore necessitates more complicated decompositions with non-zero commutators taken into account.
	
	\subsection{Spline fitting}
\label{subsec:spline_fitting}
	
	\begin{figure*}[ht]
		\centering
		\includegraphics[width=.3\textwidth,page=57]{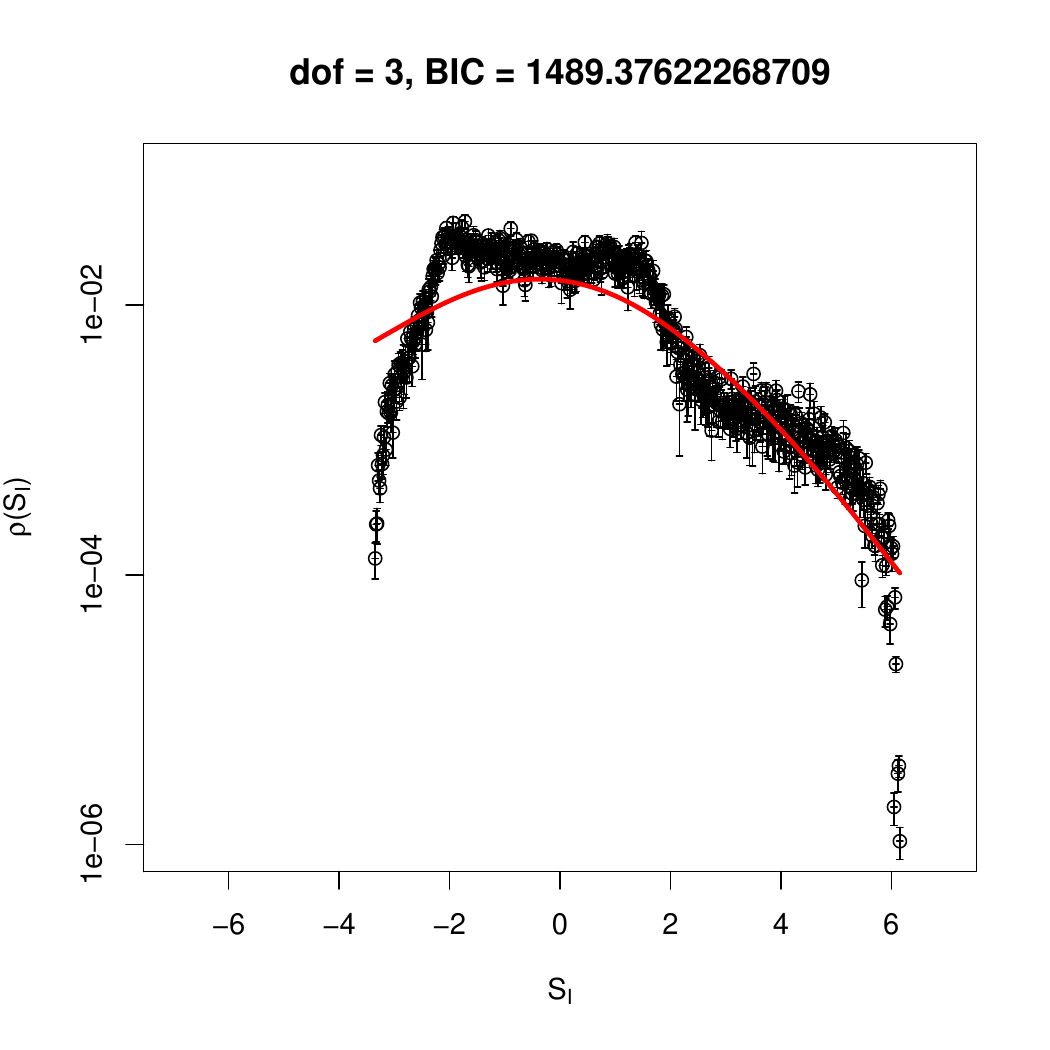}
		\includegraphics[width=.3\textwidth,page=60]{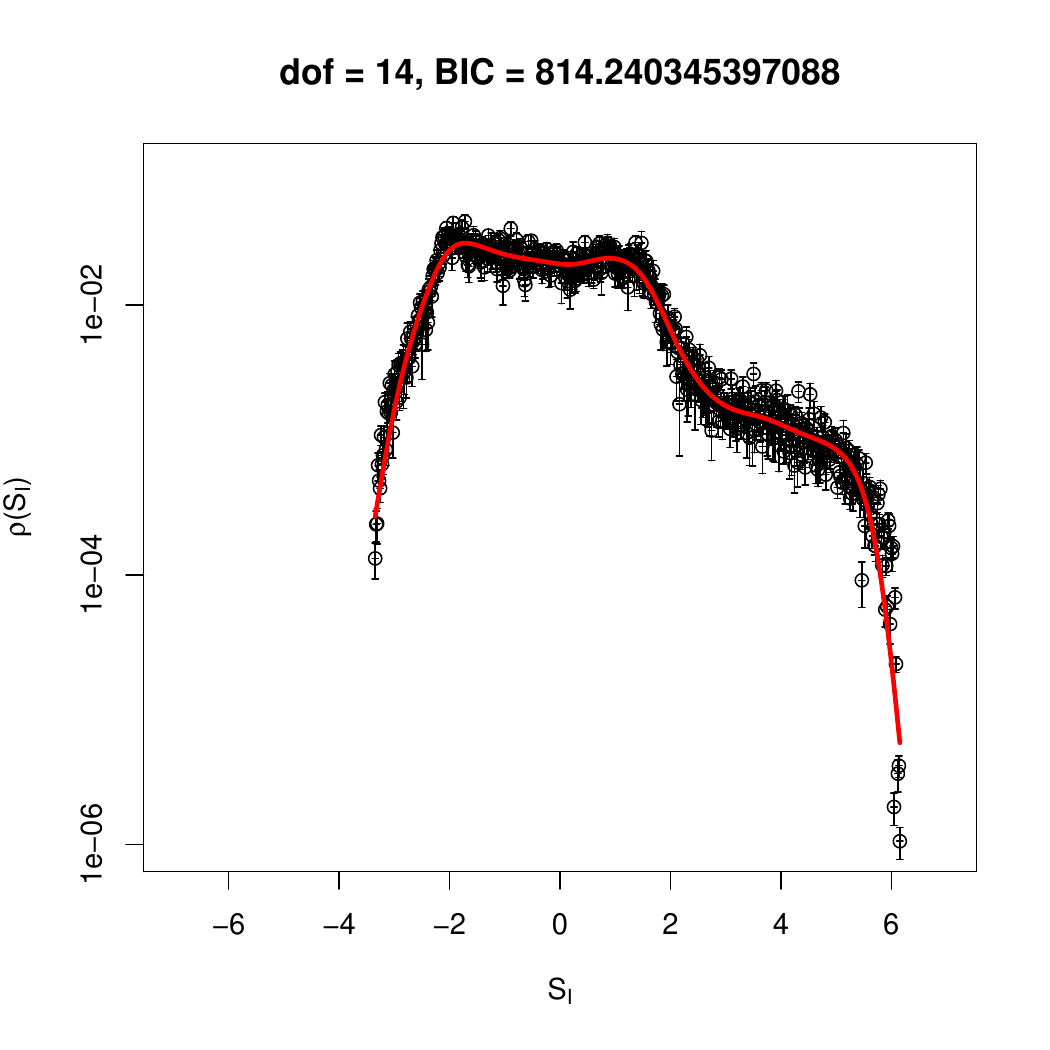}
		\includegraphics[width=.3\textwidth,page=64]{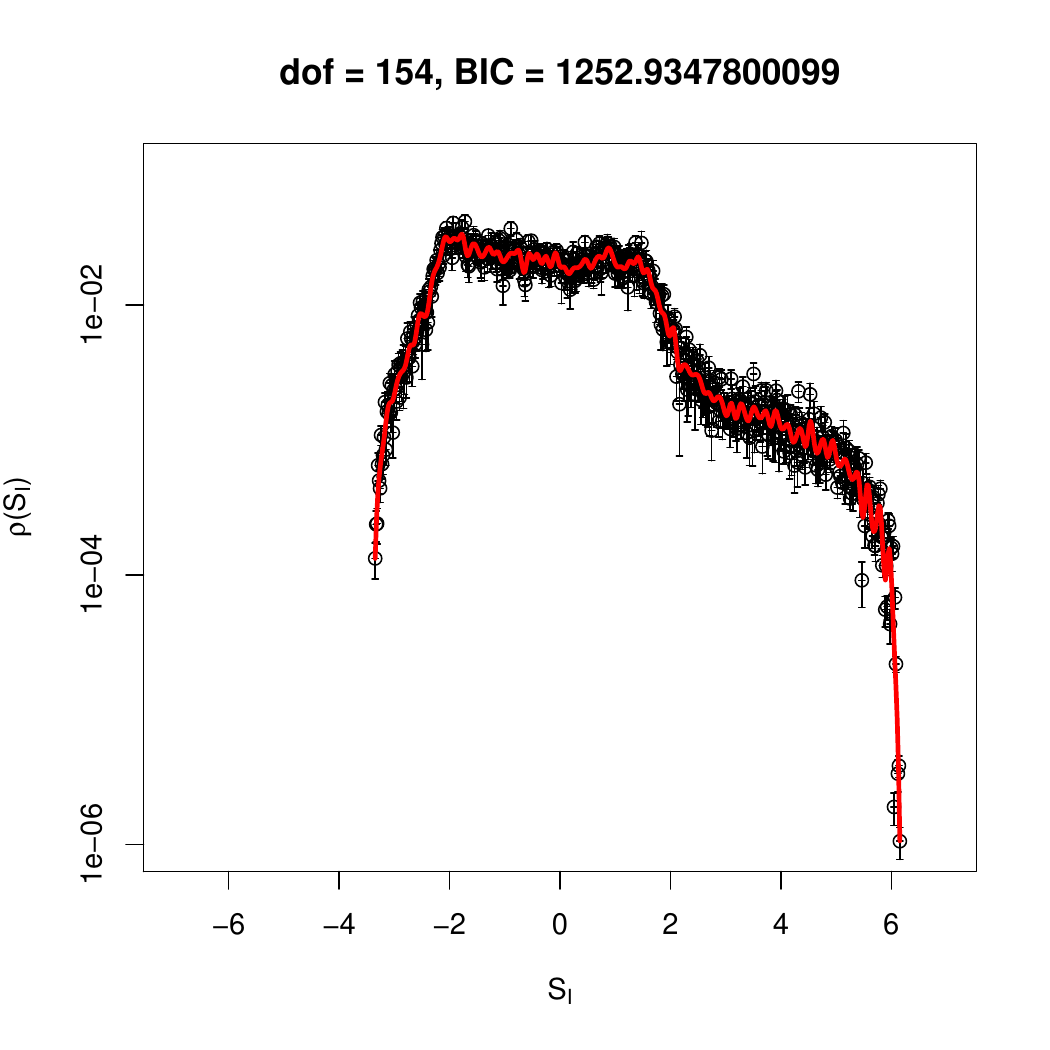}
		\caption{Spline fits to $\log\left(\rho_+(S_I) - \rho_-(S_I)\right)$ with different degrees of freedom (d.o.f.) for 538 data points and respective Bayesian information criterion (BIC), eq.~\eqref{eq:bic}. The spin chain length is $L=6$, the evolution time is $t=1$, the number of Monte-Carlo sweeps is $N_s=10^4$.}\label{fig:splines}
	\end{figure*}

	In LLR$_2$ simulations all the leading peaks have been removed from the DoS so that the remaining function is relatively smooth and can be approximated by a fit. Since we do not know anything about the global structure of the DoS in general, a smooth local approximation like a cubic spline\footnote{We use the \texttt{smooth.spline} implementation in \texttt{R}~\cite{r_language}.} is the canonical choice. Such a spline with a fixed number of free parameters or degrees of freedom (d.o.f.), where d.o.f.\ smaller or equal the number of data points, is fitted to the log-DoS as depicted in figure~\ref{fig:splines}. The fit can then be numerically integrated to high precision.
	
	A smooth fit comes with two advantages. First, discretisation errors in $S_I$ due to a finite bin size in (\ref{eq:dos_to_bins}) are reduced. In practice this effect is usually negligible and if it is not, a finer discretisation should be chosen in the first place. More relevantly, fluctuations of neighbouring bins with large uncertainties are averaged out. This allows decent estimations of the SFF at relatively high noise levels.
	
	The single but crucial disadvantage of fitting the data is that a priori the optimal number $k_\text{dof}$ of d.o.f.\ is completely unclear. In figure~\ref{fig:splines} we show the same data fitted with different order splines to the effect that the leftmost is under-fitted and will yield completely wrong results, whereas the rightmost is over-fitted and the desired smoothing effect is absent. We use the Bayesian information criterion (BIC)~\cite{bayesian_measures}
	\begin{align}
		\text{BIC} &= M\ln\left(\chi^2/M\right) + k_\text{dof} \ln (M)\label{eq:bic}
	\end{align}
	to find the optimal number of d.o.f. Here $M$ is again the number of bins of the DoS and $\chi^2$ is the usual chi-squared value. The BIC is supposed to be minimal for the best choice of $k_\text{dof}$.

	Overall the fitting allows to extract results from otherwise too noisy data, but they come at the cost of additional uncertainty and possible bias so that high-precision data is better evaluated without any fitting. This becomes clear from the comparison of the two top plots on Fig.~\ref{fig:errors}, where the errors obtained from spline-fitted results are only considerably smaller for small numbers of Monte-Carlo updates.
	
	\subsection{Origin of the non-smooth DoS}
    \label{subsec:non_smooth_dos}

	The highly irregular shape of the DoS is the crucial difference to the DoS of models with continuous (and compact) variables where the LLR approach has been shown to outperform reweighting significantly~\cite{Garron:1703.04649}. It is therefore important to understand the origin of this shape. Na\"ively one might think that it has to do with the random coupling and the chaotic nature of the system, but it turns out that this is not the case. Even the simplest transverse Ising model with constant coupling and no next-to-nearest neighbour interaction features a non-smooth DoS.
	
\begin{figure}[ht]
		\centering
		\includegraphics[width=.45\textwidth]{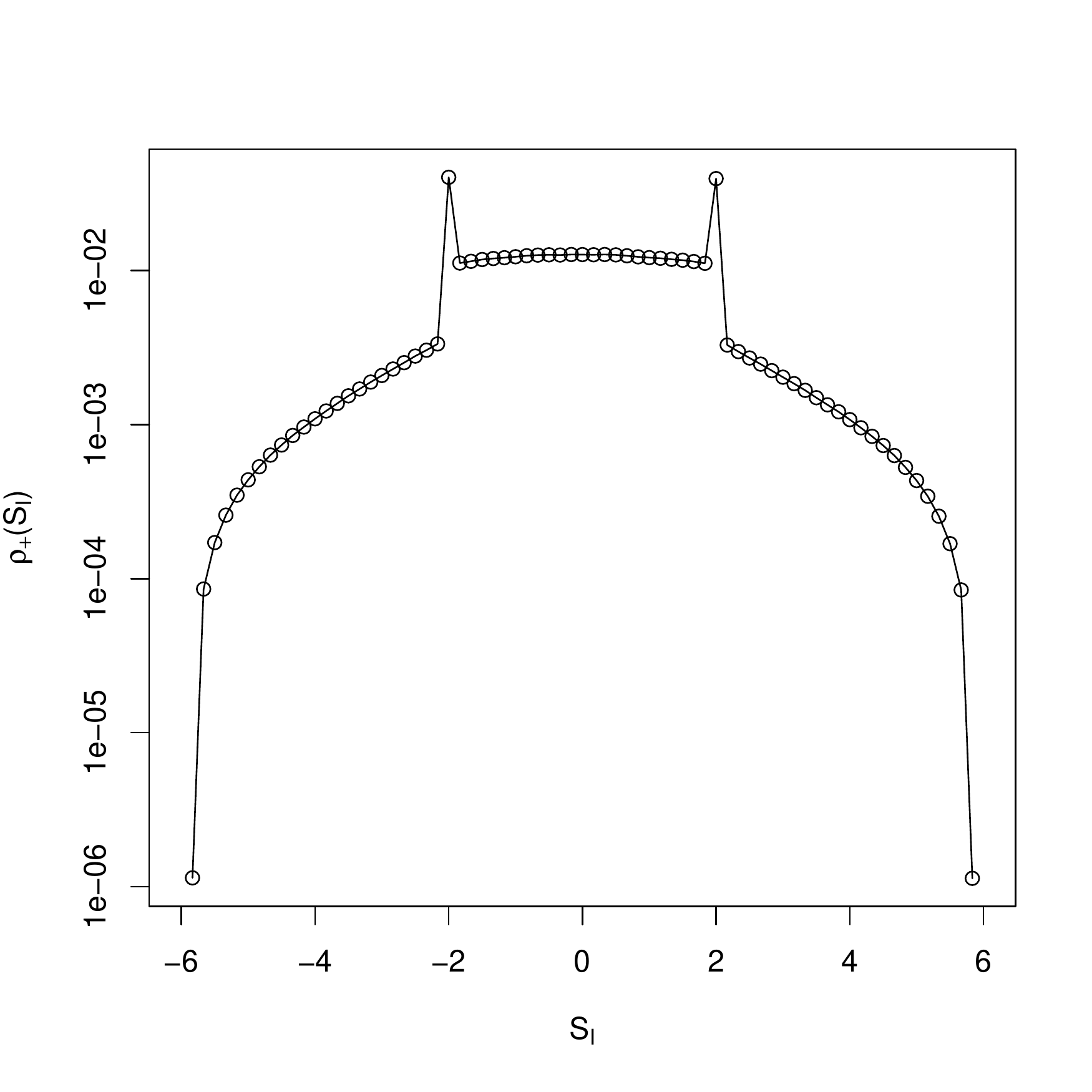}
		\caption{Density of states in the even (positive) parity sector, calculated with LLR$_2$ for the pure transverse Ising model with constant coupling, i.e.\ $\mJ_2=\Delta \mJ=0$. Chain length $L=6$; time, field strength and coupling $t=\mh=\mJ_0=1$; number of Monte-Carlo sweeps $N_s=10^7$.}\label{fig:batman}
	\end{figure}

	Qualitatively the origin of the DoS's strange form can be understood starting with the spikes in LLR$_0$ forming the highest `band'. In the case when the nearest-neighbor couplings contain quenched disorder $\Delta \mJ_i$, there are $2^{L-1}$ spikes of equal height. In the case of constant couplings, for example, for the conventional transverse field quantum Ising model without quenched disorder, there are $\lfloor L/2\rfloor+1$ spikes with binomially distributed heights. Now the second band, i.e.\ the single flip-pair expansion, can be seen (up to a proportionality factor) as a convolution of the first band with a rectangular function, taking the value 1 for every $S_I$ that can be reached by a single flip-pair and the value 0 everywhere else. Thus every delta-peak smears out to a superposition of rectangle-shaped functions at the same position, as can be seen in equation~\eqref{eq:cont_rho_n1}. Its width relative to the total range of $S_I$ scales as $\ordnung{1/L}$. The generalisation is straight forward. With every additional flip-pair the previous distribution is convoluted with a rectangular function. For instance the DoS of LLR$_2$ looks like the original spikes of LLR$_0$ convoluted with a triangle function, yielding the `batman' profile in figure~\ref{fig:batman} for the simple transverse field Ising model with constant coupling $\mJ_{ij} = \delta_{i,j+1}$ and $L=6$.
	
	Note that the notion of successive convolutions is only exact up to a single flip-pair. Beyond, it still allows to understand the shape of the DoS qualitatively but crucial contributions from interferences between two or more flip pairs are neglected (e.g.\ two flip pairs might be on the same spatial site) so that the DoS in not reproduced correctly.
	
	With larger $L$ the DoS smooths out in the disordered case. The overall smoothing stems from an exponentially large number of peaks constituting LLR$_0$, not from a broader smearing of the lower bands, since the convolution always acts on scales of $\ordnung{1/L}$.  	%
	\section{Runtime comparison}\label{sec:runtimes}
	\subsection{Exact diagonalisation}
	The straight forward solution for arbitrary times is the exact diagonalisation of the full $2^L$-dimensional system. Since the diagonalisation of a matrix has cubic scaling, we end up with a runtime in $\ordnung{2^{3L}}$.
	
	For completeness we also mention that the brute force approach of simply trying all the combinations in the classical system to calculate the exact partition sum has a runtime of $\ordnung{2^{LN_t}}$.
	
	\subsection{Reweighting}
	Stochastic methods like reweighting cannot be compared to exact methods like ED directly since the computational effort crucially depends on the desired precision $\varepsilon$. It is, however, reasonable to expect square-root convergence in the amount of statistics due to the central limit theorem and to set a sweep over the full lattice of $N = L \, N_t$ sites as the smallest unit of data. With these assumptions we get a runtime scaling of $\ordnung{LN_t\varepsilon^{-2}}$ for a system without sign problem. It is important to keep in mind that this is the best case scenario and phenomena like critical slowing down are completely neglected.
	
	For the sake of simplicity we are going to assume a small enough and constant Trotter step size $\delta$, so that $N_t\propto t$. Note that this implies exponential runtime even without sign problem for physically interesting long times $t\sim 2^L$.
	
	The crucial change we have to introduce for systems with sign problem is a factor of the statistical power
	\begin{align}
		\Sigma &= \left|\frac{Z}{Z_R}\right|\,,
	\end{align}
	that is the modulus of the expectation value of the complex phase. This modification leads to a runtime scaling of
	\begin{align}
	 T_\text{sim} \sim 	\ordnung{Lt \varepsilon^{-2} \Sigma^{-2}}\,, \label{eq:reweighting_complexity1}
	\end{align}
	so in the following we have to estimate the magnitude of $\Sigma$ as best we can. It is clear that we cannot calculate it exactly since that would allow to solve the complete initial quantum mechanical problem.
	
	We start with the observation that	
	\begin{align}
		\left|\tr U(t)\right| &= \left|A Z_R\right|\, \Sigma\,,
	\end{align}
	where the left hand side can be estimated from universal properties of the spectral form factor (SFF) and the product $A Z_R$ is known analytically. Let us expand the latter product near the continuum limit
	\begin{widetext}
	\begin{align}
		\left|A Z_R\right| &= \prod_i \left(\sin\delta \mh_i\cos\delta \mh_i\right)^{N_t/2}\left(\cosh\left(-\frac 12 \ln\tan\delta \mh_i\right)^{N_t}+\sinh\left(-\frac 12 \ln\tan\delta \mh_i\right)^{N_t}\right)\\
		&= \prod_i \left(\delta \mh_i\right)^{N_t/2}\left(\left(\frac{1}{\sqrt{\delta \mh_i}}+\sqrt{\delta \mh_i}\right)^{N_t}+\left(\frac{1}{\sqrt{\delta \mh_i}}-\sqrt{\delta \mh_i}\right)^{N_t}\right)\left(1+\ordnung{\delta \mh_i}\right)\\
		&= \prod_i\left(\left(1+\delta \mh_i\right)^{N_t} + \left(1-\delta \mh_i\right)^{N_t}\right)\left(1+\ordnung{\delta \mh_i}\right)\\
		&= \prod_i \left(\left(1+\frac{t \mh_i}{N_t}\right)^{N_t}+\left(1-\frac{t \mh_i}{N_t}\right)^{N_t}\right)\left(1+\ordnung{\delta \mh_i}\right)\\
		&= \prod_i 2\cosh\left(t \mh_i\right)\left(1+\ordnung{\delta \mh_i}\right)\,.
	\end{align}
	\end{widetext}
	In the common case of all fields equal $\mh_i\equiv \mh$ we thus obtain the formula
	\begin{align}
		\Sigma &= \frac{\left|\tr U(t)\right|}{2^L\cosh\left(t \mh\right)^L}\,,
	\end{align}
	which is exact as $\delta \rightarrow0$.
	
	Since the SFF is bound from above $\left|\tr U(t)\right|\le 2^L$, the best possible runtime scaling amounts to
	\begin{align}
		T_\text{sim} \sim \ordnung{Lt\varepsilon^{-2}\cosh\left(t \mh\right)^{2L}}\,,\label{eq:best_reweighting_scaling}
	\end{align}
	so that reweighting can never outperform exact diagonalisation when $t\mh\gtrsim \acosh 2\approx \num{1.3}$.
	
	Empirically we find that for small times $\left|\tr U(t)\right|\sim 2^L\eto{-Lt^2}$ until it reaches the plateau at $2^{L/2}$. Moreover, in order to maintain a high precision, usually we have to scale the Trotter step with the system size $\delta\sim1/L$, so that a more realistic runtime estimate is given by
	\begin{align}
		T_\text{sim} \sim \ordnung{\frac{1}{2^{-L}+\eto{-Lt^2}}\,L^2t\varepsilon^{-2}\cosh\left(t \mh\right)^{2L}}\,.\label{eq:realistic_reweighting_scaling}
	\end{align}
	
	\subsection{LLR}\label{sec:llr_eq_rew}
	In what follows we show that for both reweighting and LLR the error is strongly dominated by the `bulk', that is the region with highest DoS. Since LLR is only advantageous in sampling the `tails' of the DoS distribution more accurately, it cannot significantly outperform reweighting. This conclusion is also confirmed by our Monte-Carlo data.
	
	For a more formal comparison we have to introduce several assumptions. First of all we demand that all the LLR parameters have been tuned properly. In particular the bin size used for LLR is small enough to have negligible discretisation errors. Such a choice is always possible though the need to tune the parameters is a serious disadvantage of LLR. Secondly, we bin the data accumulated for reweighting in the same fashion, again without significant loss of precision. Now the normalised DoS obtained by either method can be interpreted as a probability distribution $\rho_i$ on the same discrete and finite set $i\in\{1,\dots,M\}$. Same as in equation~\eqref{eq:dos_to_bins}, $M$ denotes the number of bins the DoS has been split into.
	
	It is easy to see that with reweighting every bin $i$ is visited $N_i=\rho_i\Lambda$ times on average, where $\Lambda$ is the total number of samples. Therefore, in the large $\Lambda$ limit, according to the central limit theorem (or naturally assuming a Poisson distribution of bin counts), the absolute uncertainty of the number of counts in said bin is $\sqrt{N_i}$ yielding an error of
	\begin{align}
		\Delta\rho_i &\equiv \frac{\rho_i}{\sqrt{N_i}}\\
		&=\sqrt{\frac{\rho_i}{\Lambda}}\,.
	\end{align}

	Since LLR is a special case of the Robbins-Monro (RM) algorithm with a non-differentiable target function, the coefficients in the exponent converge strictly as $\ordnung{1/\sqrt{\Lambda}}$~\cite{nemirovskii1983problem}. Thus, the relative error of $\rho_i$ lies in $\ordnung{\eto{1/\sqrt{\Lambda}}-1}=\ordnung{1/\sqrt{\Lambda}}$. We can obtain some bounds on the proportionality factor as well, though these results are by no means as precise as for reweighting.
	
	In the early stages of the RM iteration the probability to update the $i$-th bin is simply $\rho_i$ (same as for reweighting). As Robins-Monro iterations proceed and the coefficients $\alpha_i$ are updated, at asymptotically large $\Lambda$ all bins are visited with equal probabilities $\sim 1/M$. Therefore in the \textit{bulk} of the DoS defined as all the regions with $\rho_i \gg 1/M$ the probability to visit bin $i$ decreases with the number of iterations $\Lambda$ and eventually approaches $1/M$ from above. We can thus provide a bounds on $N_i$ in the bulk region:
	\begin{align}
		\rho_i \Lambda \ge N_i \ge \frac{\Lambda}{M}\,.
	\end{align}

	On the other hand, in the \textit{tail} of the distribution where $\rho_i \ll 1/M$, the probability to visit the bin $i$ increases with the number of RM iterations, and eventually approaches $1/M$ from below. The expectation value of the iteration count $N_i$ in this region is therefore bounded by
	\begin{align}
		\rho_i \Lambda \le N_i \le \frac{\Lambda}{M}\,.
	\end{align}
	In the following we assume the best possible scenario, using the upper bounds on $N_i$, in which case the statistical error is given by
	\begin{align}
		\Delta \rho_i &= \begin{cases}
			\sqrt{\frac{\rho_i}{\Lambda}} & \text{if $i$ in bulk,}\\
			\rho_i\sqrt{\frac M\Lambda} & \text{if $i$ in tail,}
		\end{cases}
	\end{align}
	and show that LLR cannot outperform reweighting significantly even in that case.
	
	Let us consider some probability $\rho_0$ from the bulk and another $\rho_\infty=\zeta\rho_0$ from the tail, where $\zeta$ is some small parameter. For reweighting we need $\Lambda\sim\zeta^{-2}$ samples to resolve $\rho_\infty$ at all. For LLR on the other hand the error in the tail is much smaller. But even if we knew $\rho_\infty$ exactly, this knowledge is useless as long as $\Delta \rho_0 > \rho_\infty$, i.e.\ the error is bulk-dominated. From this follows the condition
	\begin{align}
		&&\Delta \rho_0 &\overset{!}{<} \rho_\infty\\
		\Rightarrow&& \sqrt{\frac{\rho_0}{\Lambda}} &< \zeta\rho_0\\
		\Leftrightarrow&& \Lambda &> \zeta^{-2}/\rho_0\,.
	\end{align}

	Thus, not only does LLR follow the same scaling as for reweighting in Eq.~(\ref{eq:reweighting_complexity1}), it even has very similar errors quantitatively since the statistical uncertainties in the bulk are almost identical for reweighting and LLR, and these uncertainties dominate in both cases.
	
	We remark that this argument is applicable independently of the specific form of the DoS and the conclusion therefore generalisable beyond its application to spin chains. Generally, complexity of a given problem is encoded in the magnitude of the parameter $\zeta$. For instance in our case of real time spin chain simulations $\zeta\sim\exp(-Lt)$ is exponentially suppressed with the system size providing an alternative perspective on the origin of the NP-hardness of the sign problem.
	
	\subsection{Augmented \texorpdfstring{LLR$_2$}{LLR2} or \texorpdfstring{REW$_2$}{REW2}}
	Removing the two leading order contributions from the stochastic calculations and evaluating them exactly, comes with an additional cost of
	\begin{align}
		\ordnung{L2^L}
	\end{align}
	as compared to ``plain vanilla'' LLR$_0$ or REW$_0$. However, it allows to achieve the same precision with a significantly lower accuracy goal since the stochastic part contributes with a reduced weight. More specifically,
	\begin{align}
		\varepsilon^{-1} &\mapsto \varepsilon^{-1}\left(1-\cosh(t\mh)^{-L}\left(1+\frac12L(t\mh)^2\right)\right)
	\end{align}
	results in an overall runtime of the approximate order
	\begin{align}
		\begin{split}
			T_\text{sim} &\sim \ord\left(L \, 2^L \vphantom{\frac{L (t\mh)^2}{2}}
			\right. \\
			&\qquad\left. + 
			Lt\varepsilon^{-2}\left(\cosh\left(t \mh\right)^{L}-1-\frac{L (t\mh)^2}{2} \right)^2 \right)\,.
		\end{split}\label{eq:best_llr2_scaling}
	\end{align}
    This estimate should be compared with (\ref{eq:best_reweighting_scaling}). We see that summation over the two leading-order contributions allows to obtain equally precise results at short times with significantly smaller statistics (i.e.\ computational effort) and to extend the stochastically calculable time region towards somewhat larger values.  	%
\section{Results}\label{sec:results}
	All the simulations presented in this work have been performed using \texttt{R} as a front- and \texttt{C} as a back-end. The complete code and most of the data is publicly available at~\cite{j_ostmeyer_2022_7164902}. Some of the data is rather large and has therefore not been published in the same way, but we are happy to provide it in case of interest.

	We tested our algorithm on chains of up to $L=50$ with LLR$_0$/REW$_0$ and up to $L=40$ with LLR$_2$/REW$_2$. Small length $L\le 16$ have been benchmarked against results from exact diagonalisation.
	We emphasise that the larger lattices with $L \gtrsim 24$ are far out of the reach of any exact calculations. For instance $L=40$ would require $\SI{16}{\tera\byte}$ of memory to even store a single complex state vector in double precision.
	We also stress that while LLR$_2$/REW$_2$ has a runtime scaling as $\ordnung{2^L}$ and therefore $L>40$ quickly becomes unfeasible, there are no such limitations for LLR$_0$/REW$_0$ and $L$ can be chosen practically arbitrarily large at the cost of somewhat reducing the evolution time $t$.
	
	In each simulation we chose a single random realisation of the coupling $J$. The disorder average over $J$ is technically straight forward and would only add a layer of uncertainties to the results obscuring the quality of the algorithms.
	
	\begin{figure*}[ht]
		\centering
		\includegraphics[width=.45\textwidth,page=1]{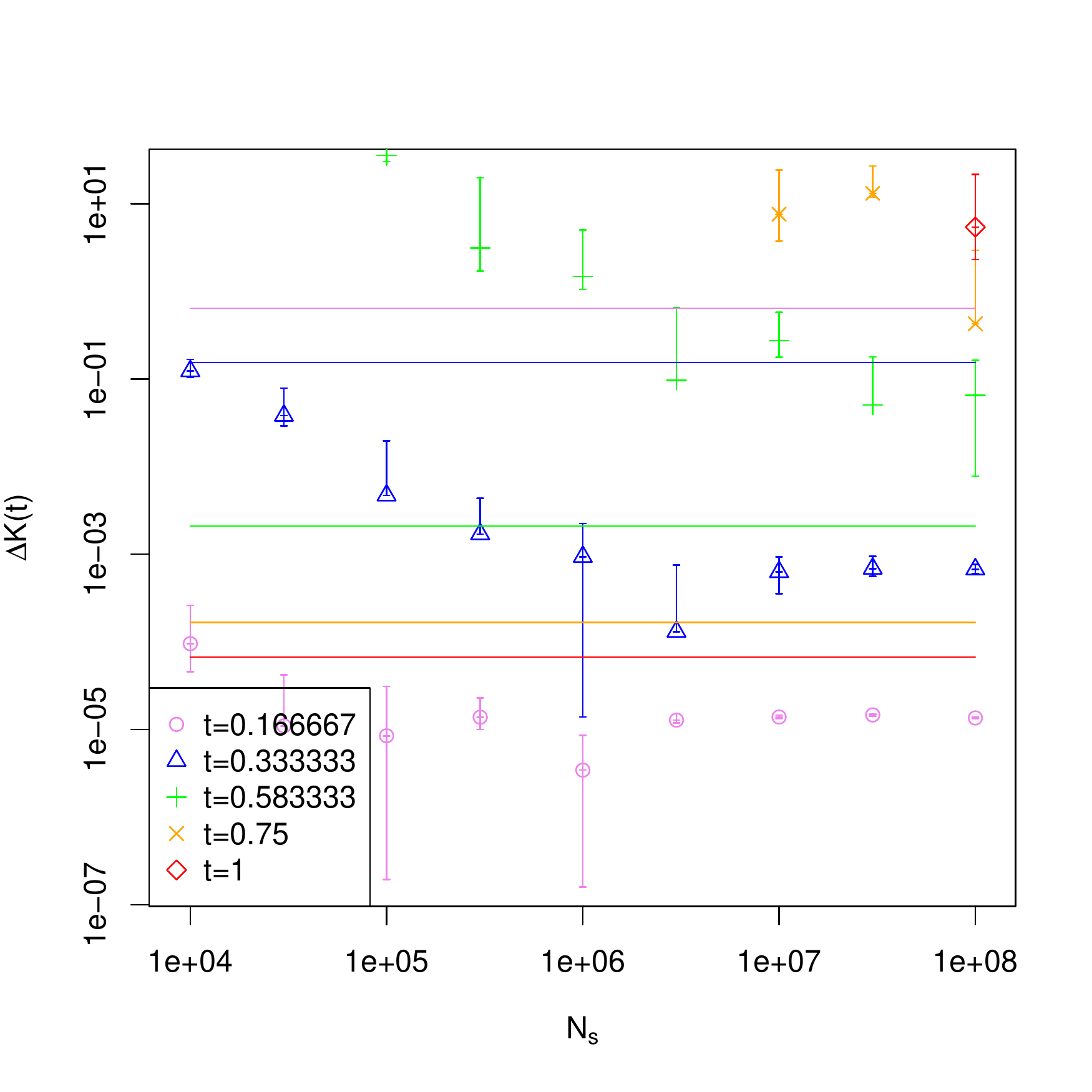}
		\includegraphics[width=.45\textwidth,page=2]{error_L_16_dJ_1_av_2.pdf}
		\includegraphics[width=.45\textwidth,page=4]{error_L_16_dJ_1_av_2.pdf}
		\includegraphics[width=.45\textwidth,page=3]{error_L_16_dJ_1_av_2.pdf}
		\caption{Relative errors of the spectral form factor $\Delta K(t) = \left|K(t)/K_\text{ED}(t)-1\right|$, where $K_\text{ED}(t)$ has been calculated with exact diagonalisation, as a function of the number of sweeps $N_s$. For small evolution times $t \lesssim 0.3$, $\Delta K\lr{t}$ reaches a plateau value that is entirely saturated by non-statistical Trotterization errors. Error bars in negative direction that would extend below 0 are omitted. Lines show the respective statistical powers. Top left to bottom right: LLR$_2$, LLR$_2$ with spline fitting, LLR$_0$, REW$_2$. The chain length is $L=16$ for all plots.}\label{fig:errors}
	\end{figure*}
	
	Some typical examples of the scaling of the relative error between the Monte-Carlo and exact diagonalization results with the number of Monte-Carlo iterations are shown on Fig.~\ref{fig:errors}. The error is plotted as a function of the total number of sweeps $N_s=\Lambda/N$ for different evolution times $t$ and correspondingly different strengths of the sign problem. Respective statistical powers $\Sigma$ are shown as lines. In an ideal stochastic simulation the error should scale as $\Sigma^{-1}/\sqrt N_s$ and it does so usually. There are some exceptions in the decrease with $N_s$ though. The shortest evolution times just plateau out at a constant error level. For LLR$_0$ the simulations would require a much finer discretisation of $S_I$ in order to resolve the spiked DoS properly. The LLR$_2$ and REW$_2$ calculations on the other hand are so precise that the Trotter error for the particular choice of $N_t$ is resolved.

	From the two top plots in Fig.~\ref{fig:errors} we also conclude that the results obtained with spline fitting tend to be better than the ones from simple integration for small numbers $N_s \lesssim 10^5$ of Monte-Carlo updates, and for large evolution times $t \sim 1$ where the sign problem is more severe. More statistics undermines the usefulness of fits as they cannot extract any additional information.
	
	The true reason for the favourable spline results in the case of small statistics appears to be that the best fitting functions are very smooth in this case. Therefore the highly oscillatory integral~\eqref{eq:partition_sum_dos} required to obtain the spectral form factor yields a value close to zero. Since the analytic approximation used in LLR$_2$ is not too bad by itself, the overall result is better than the noise in the unfitted case. Thus, using the stochastic data does not help to improve the analytic approximation for too little statistics. But in the spline-fitted case it does not add any harm either.

	It is evident that the augmentation defining LLR$_2$ allows to reduce the error at short times by several orders of magnitude compared to ``plain vanilla'' LLR$_0$. It is also clear, however, that it does not significantly mitigate the sign problem at long times confirming the predictions from equation~\eqref{eq:best_llr2_scaling}.
	
	A direct comparison of LLR$_2$ and REW$_2$ shows very little differences. If anything, reweighting appears to be slightly better. This is exactly what we expect from \cref{sec:llr_eq_rew} where we showed that LLR and reweighting have principally the same sign problem. The minor difference in favour of reweighting most likely comes from the binning required for LLR as well as a suboptimal parameter choice (see \cref{sec:llr_params}).

	\begin{figure*}[ht]
		\centering
		\includegraphics[width=.45\textwidth,page=1]{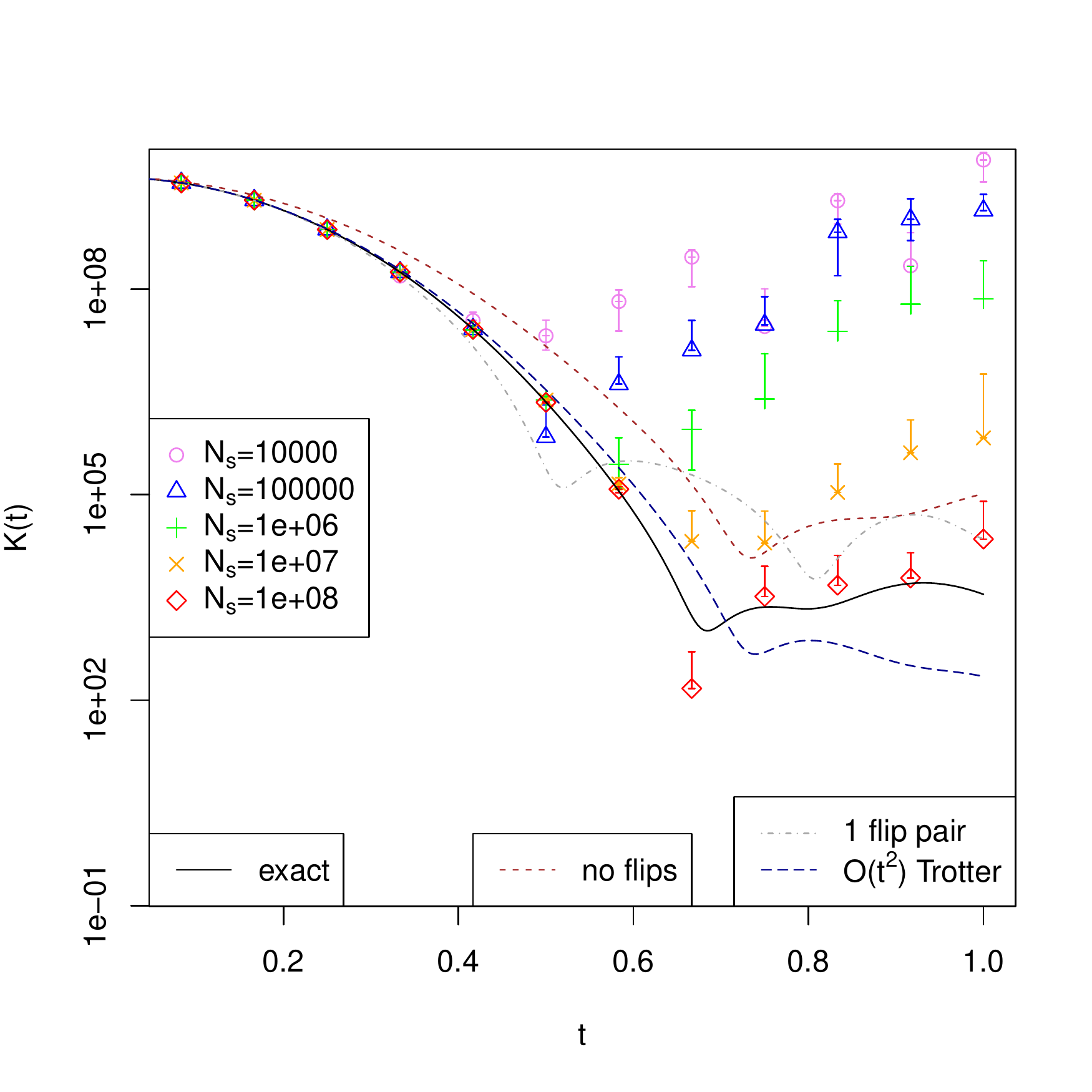}
		\includegraphics[width=.45\textwidth,page=2]{sff_llr_L_16_dJ_1_av_2.pdf}
		\includegraphics[width=.45\textwidth]{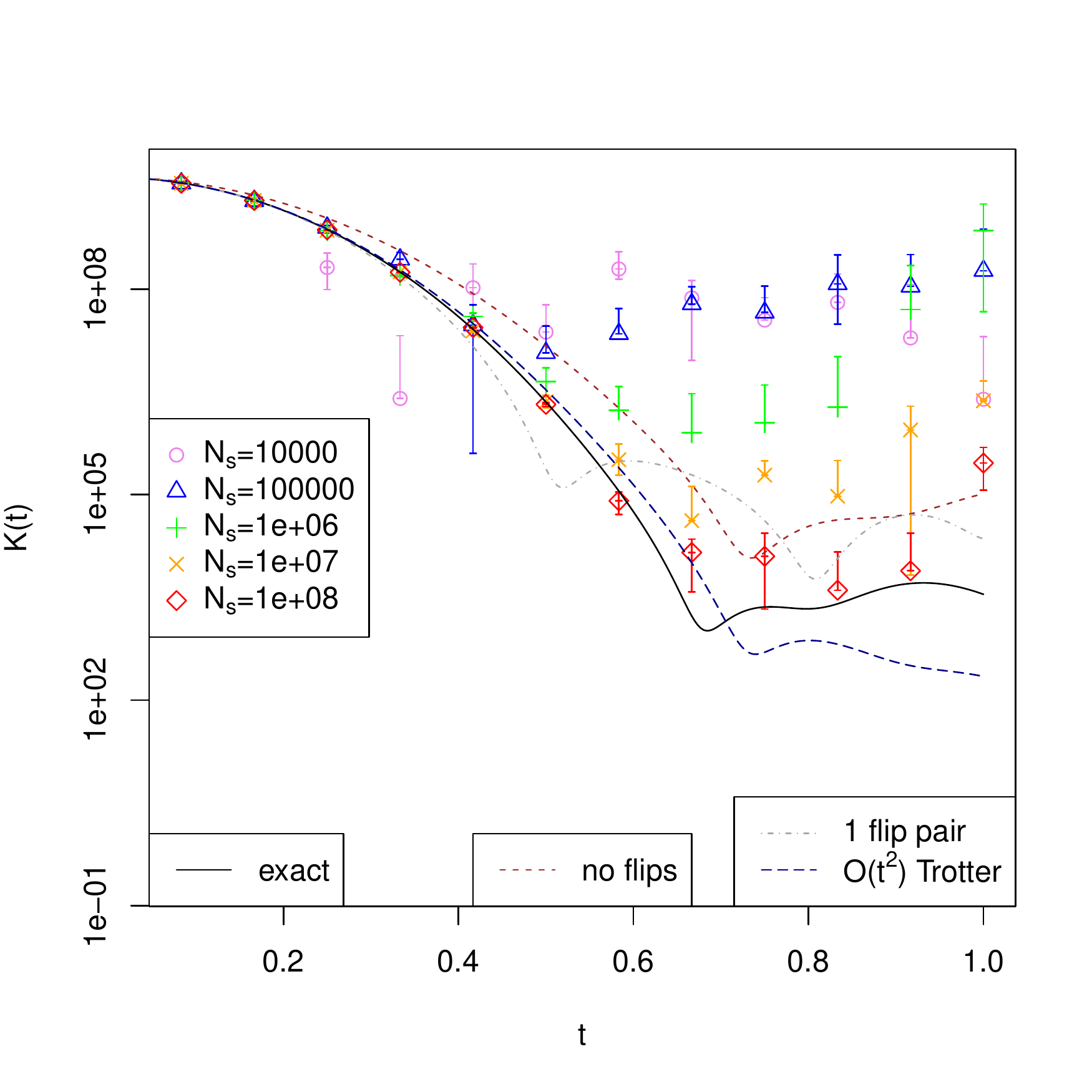}
		\includegraphics[width=.45\textwidth]{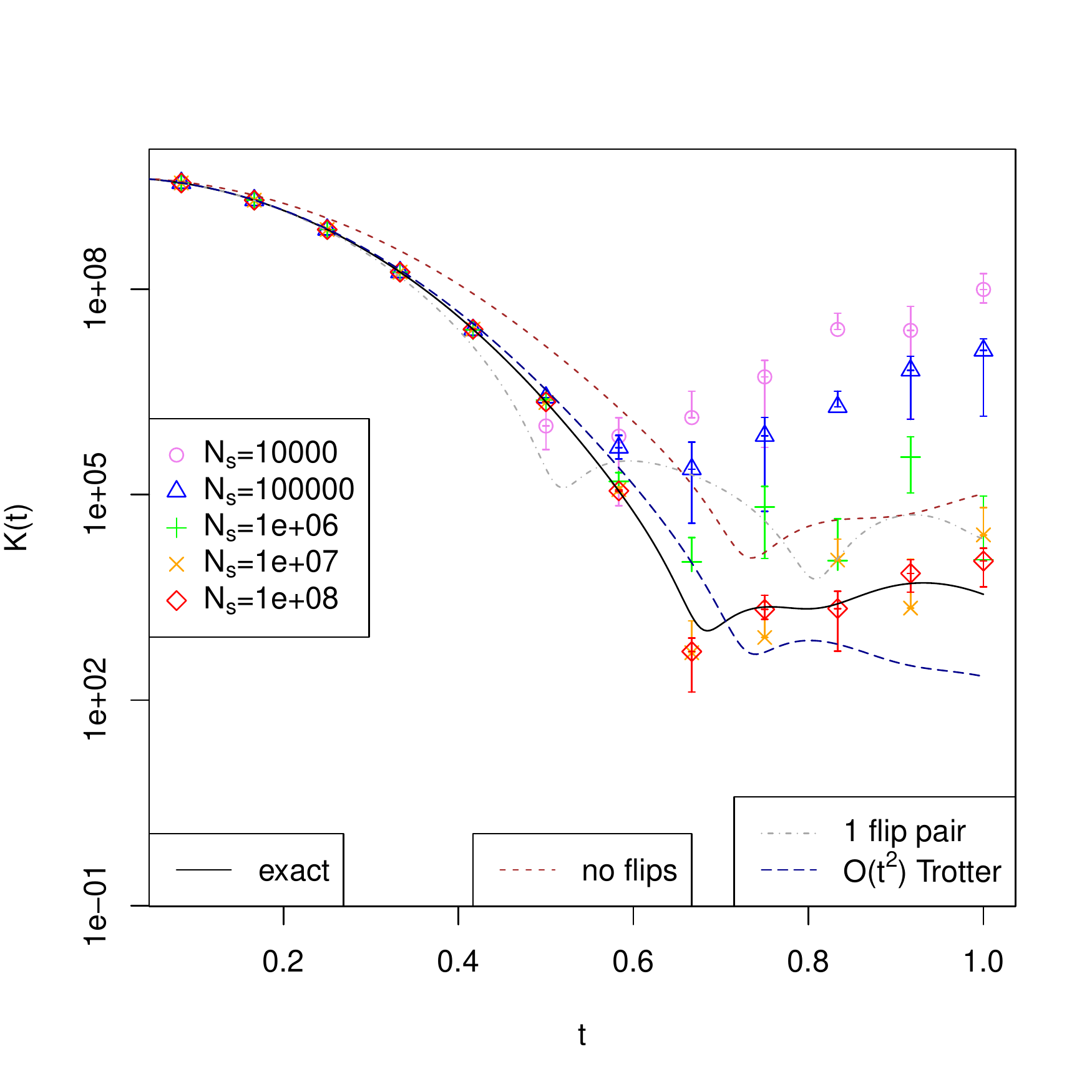}
		\caption{The spectral form factors $K(t)$ compared with the solution from exact diagonalisation (``exact'') and the approximations ``no flips''~\eqref{eq:no_flip}, ``1 flip pair''~\eqref{eq:single_flip}, and ``$\ordnung{t^2}$ Trotter''~\eqref{eq:first_order_trotter} for different numbers of Monte-Carlo sweeps $N_s$. Error bars in negative direction that would extend below 0 are omitted. Top left to bottom right: LLR$_2$, LLR$_2$ with spline fitting, LLR$_0$, REW$_2$. The chain length is $L=16$ for all plots.}\label{fig:form_factors}
	\end{figure*}

	Figure~\ref{fig:form_factors} shows the spectral form factors corresponding to the errors in figure~\ref{fig:errors}. We added the approximations for continuous times derived in \cref{sec:approx_formulae}. The strength of the LLR$_2$ method becomes clear when the small differences between the approximations and the exact results are observed. Only these differences have to be calculated stochastically. Note, however, that even at very short times $t\lesssim\num{0.4}$ the stochastic contribution cannot be neglected and the Monte-Carlo results are not compatible with the approximations.
	
	We further remark that the LLR results at large times and small statistics are systematically incompatible with the correct values. This can be explained by a required minimal number of iterations to `fill up' the DoS. We expect that a different choice of coefficients $a_{L,M,\Lambda}$ and $b_M$ than in equations~\eqref{eq:opt_a} and~\eqref{eq:opt_b} might reduce this problem, but we did not conduct additional extensive parameter tuning. This is certainly a great disadvantage of LLR as opposed to reweighting.
	
	\begin{figure*}[ht]
		\centering
		\includegraphics[width=.45\textwidth]{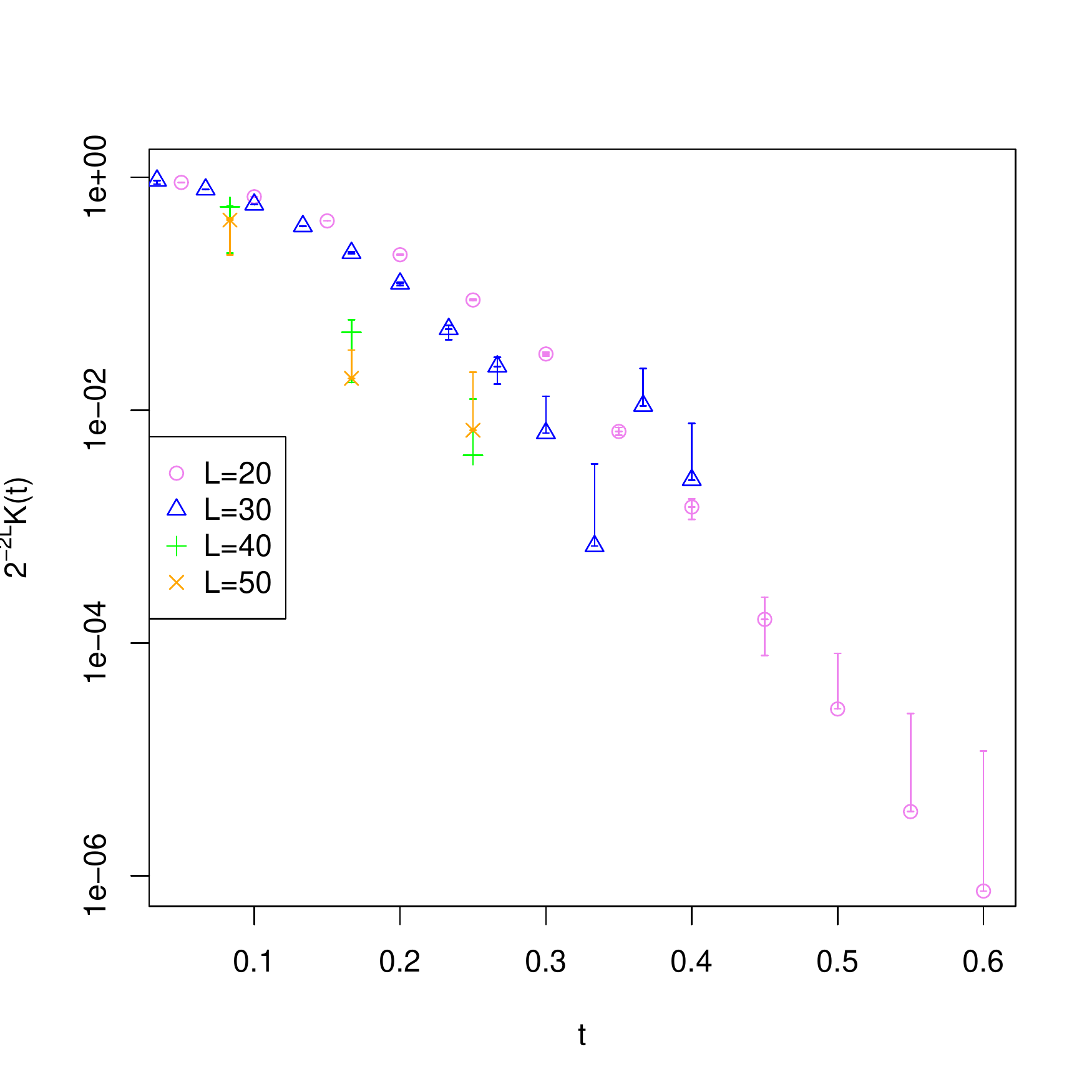}
		\includegraphics[width=.45\textwidth]{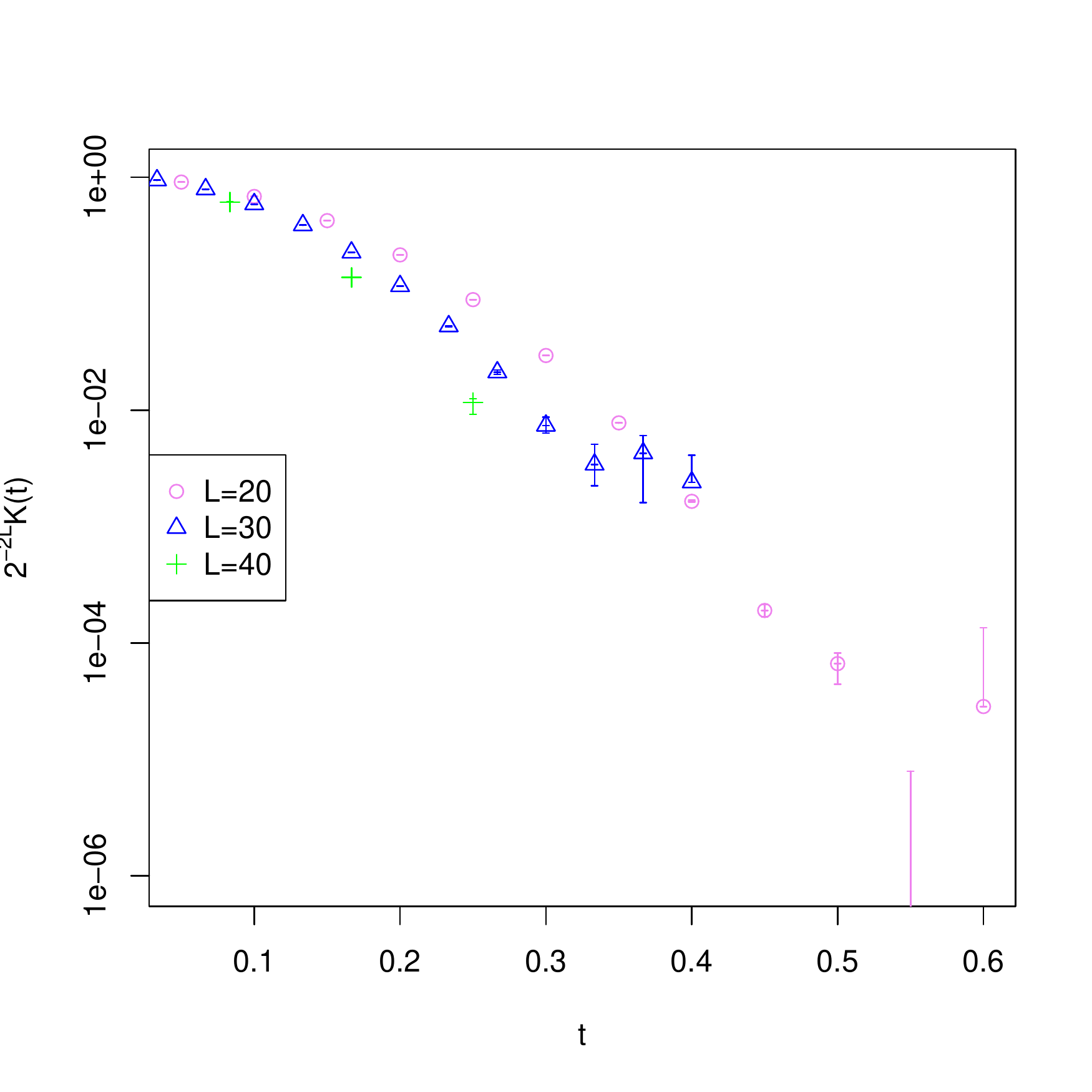}
		\includegraphics[width=.45\textwidth]{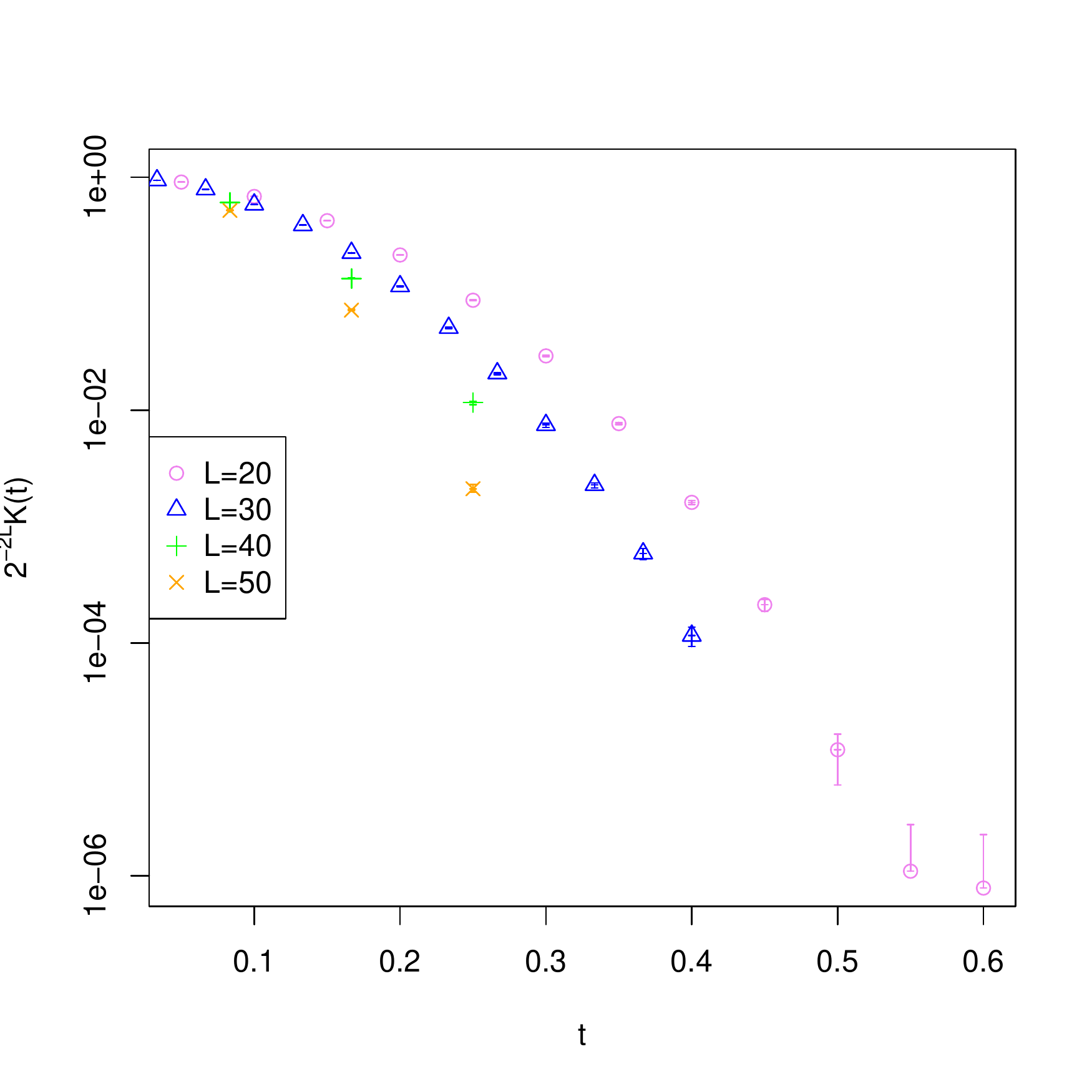}
		\includegraphics[width=.45\textwidth]{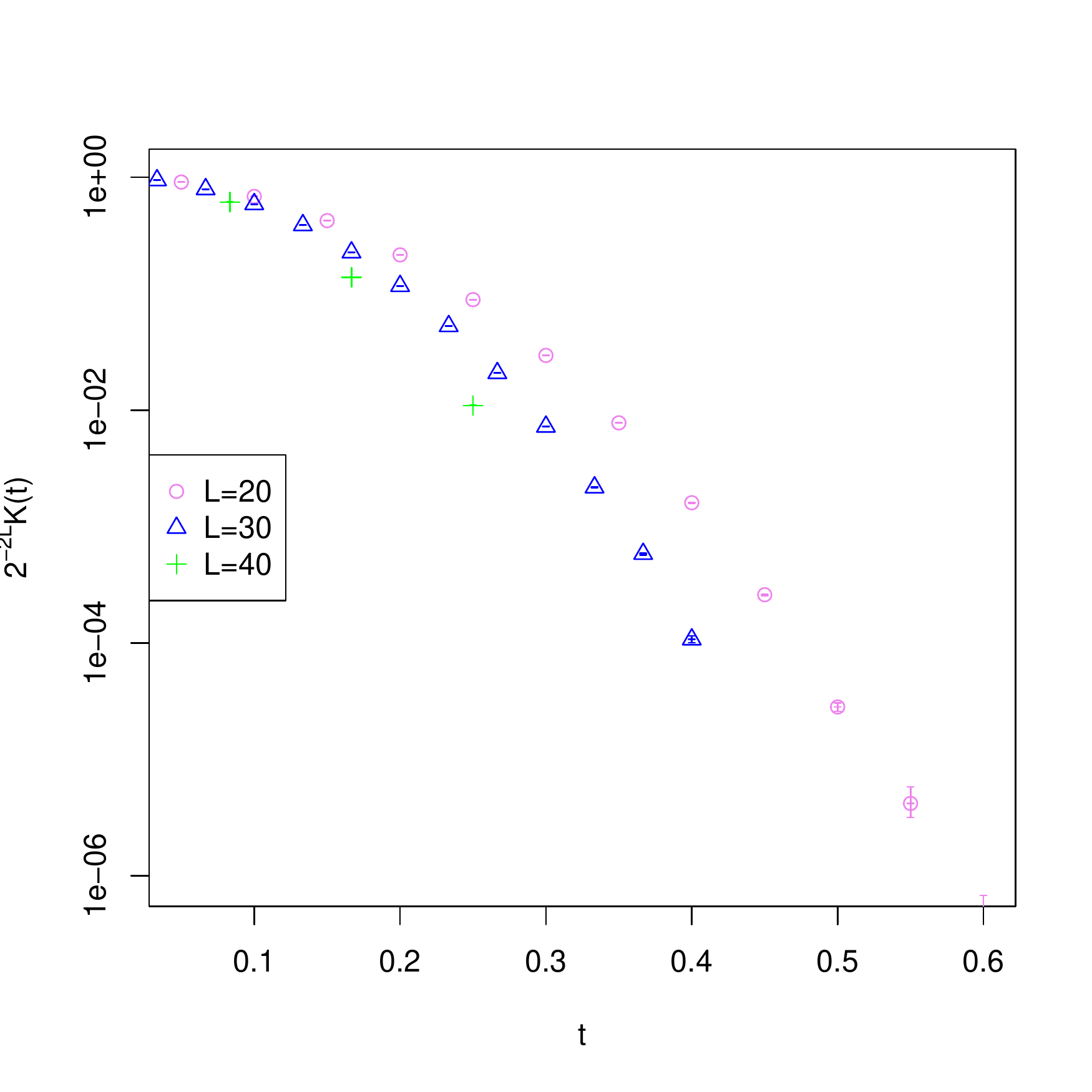}
		\caption{Normalised spectral form factors $2^{-2L} K(t)$ for system sizes hardly ($L=20$) or not at all ($L\ge30$) treatable by exact diagonalisation. Error bars in negative direction that would extend below 0 are omitted. Top left to bottom right: LLR$_0$, LLR$_2$, REW$_0$, REW$_2$. The number of Monte-Carlo sweeps is $N_s=10^7$ for all plots.}\label{fig:sff_large}
	\end{figure*}

	The results obtained for longer chains $L\ge20$ and depicted in figure~\ref{fig:sff_large} are qualitatively similar to those discussed above for $L=16$. Again LLR$_2$ and REW$_2$ yield extremely precise results at short times and reweighting generally outperforms LLR. Notably these results demonstrate the scalability of our stochastic algorithms towards very long chains provided $t\lesssim1$. Especially for REW$_0$ we can obtain reliable results for virtually arbitrarily large values of $L$.  	%
\section{Discussion and outlook}
\label{sec:outlook}

    In this work, we investigated the possibility to evaluate the spectral form-factor (real-time partition sum) of a non-integrable spin chain (\ref{Hamiltonian}) using Monte-Carlo simulations. The partition sum contains a multitude of mutually cancelling complex-valued contributions, which turns its direct evaluation into an NP-hard problem. We investigated the infinite-temperature spectral function because it poses the most challenging problem and is also often considered in the literature. We remark however that the method would be applicable with only minor modifications (a larger, in principle unlimited, number of parity sectors) to finite temperature systems as well. In fact, finite temperatures might severely reduce the sign problem.

    We compared the most straightforward approach based on importance sampling with subsequent reweighting with a more advanced Density-of-States (DoS) approach, in which the statistical distribution of the complex phase of the action is calculated with high precision.

    One of our main results is the improved simulation strategy, in which the contribution of configurations with zero and one spin flips into the real-time partition function are summed over exactly. These configurations contribute to the leading and the next-to-leading orders of the expansion of the real-time partition function $\tr U\lr{t}$ in (\ref{spf_def}) in powers of the magnetic field $\mh$. Summation over all other configurations is performed stochastically, using either the standard Monte-Carlo updates (for reweighting), or the LLR algorithm (for the DoS approach). As we discuss below, both approaches work nearly equally well.

    As illustrated in Fig.~\ref{fig:errors}, the improved simulation strategy significantly accelerates convergence to exact results, which we obtained using numerical diagonalization of the Hamiltonian. It allows to study real-time evolution of spin chains with lengths up to $L = 40$ and up to physical times $t \lesssim 1$. Our method thus clearly outperforms numerical exact diagonalization methods at short evolution times $t \lesssim 1$. Since for our spin chain the numerical cost of exact diagonalization does not depend on the evolution times (up to possible issues with numerical precision of the result due to round-off errors), exact diagonalization becomes more advantageous for $t \gtrsim 1$. The algorithm could also be translated to a continuous time formulation along the lines of~\cite{PhysRevLett.77.5130}, which removes trotterization errors.

    An advantage of the stochastic approach is that it can be easily extended to bosonic systems with infinite-dimensional Hilbert spaces, for which exact diagonalization typically becomes prohibitively expensive even for $O\lr{10}$ degrees of freedom. Indeed, to obtain precise results for bosonic degrees of freedom, the size of the local Hilbert space $N_\text{loc}$ associated with each bosonic degree of freedom should usually be much larger than $2$, which results in a significantly faster growth of computational cost $T_\text{sim} \sim \lr{N_\text{loc}}^{n_\text{dof}}$ with the number of degrees of freedom than for quantum spin chains. In addition, we need to extrapolate the result to the limit $N_\text{loc} \rightarrow +\infty$. The possibility to simulate short periods of real-time evolution for systems with a large number of bosonic degrees of freedom is particularly attractive for numerical studies of the early-time evolution of quark-gluon plasma in heavy-ion collisions before the onset of hydrodynamic behavior.

    Another important and somewhat counter-intuitive conclusion is that, at least for our particular Hamiltonian and the trotterization scheme, the DoS/LLR approach does not appear to offer significant advantage over the simple reweighting. In fact, reweighting often outperforms the DoS/LLR approach, presumably, because of the sub-optimal choice of parameters in the LLR algorithm. Reweighting, on the other hand, does not depend on any tunable parameters.

    In general, the main technical advantage of the DoS/LLR approach is that it allows to resolve the tails of the distribution of the complex phase with much higher precision than reweighting. In our case, the contribution of these tails to the real-time partition sum is sub-dominant in comparison to the bulk contribution (central region in the distributions on Fig.~\ref{fig:dos_llr0_vs_llr2}). Therefore, sampling the tails with high precision is not an advantage in itself. As stressed in \cite{Langfeld:1404.7187,Langfeld:1605.02709,Garron:1703.04649,Schnack+2021}, high-precision data for the tails of the density of states becomes an advantage if the density of states $\rho\lr{E}$ can be well approximated by some function with a small number of parameters, e.g.\ a low-degree polynomial fit or a spline. In this case, precise knowledge on the distribution tails translates into a better choice of fit parameters. This allows for more precise numerical estimates of the oscillatory integrals of the form $Z = \int \md E \rho\lr{E} \eto{\im E}$ in the representation (\ref{path_int3}) of the real-time partition function.

    As discussed in Section~\ref{subsec:non_smooth_dos}, in our case the DoS in Eq.~(\ref{eq:dos_definition}) is a sum of $\delta$-functions and piecewise constant terms for any finite $L$. The DoS only looks like a continuous function in the thermodynamic limit $L \rightarrow +\infty$. While separating out the leading and the next-to-leading contributions to the DoS, as described in Subsections~\ref{subsec:leading_order}~and~\ref{subsec:next_to_leading_order}, leaves us with a much smoother DoS, we were still not able to approximate it by a function (polynomial or spline) with a sufficiently small number of parameters. As a result, spline or polynomial approximations improve the convergence to exact results only for small numbers of Monte-Carlo samples in Robins-Monro iterations (\ref{eq:robins_monro}). Consequently, the LLR approach did not offer significant improvement over the reweighting approach.

    In this respect, the LLR approach might work better for systems with continuous degrees of freedom, where the Density of States $\rho\lr{E}$ in (\ref{path_int2}) is a continuous function even for a finite number of degrees of freedom (away from the thermodynamic limit). In this case, the DoS might allow for good-quality approximations in terms of functions with a small number of parameters, which allows to effectively use the information about the tails of the DoS to improve the precision of oscillatory integrals of the form (\ref{path_int3}). However, other complications might arise when applying the DoS/LLR method to continuous degrees of freedom. To provide the most obvious example, the DoS will be clearly a non-normalizable and unbounded function for the real-time partition function of even the simplest one-dimensional quantum harmonic oscillator. For the harmonic oscillator, this unboundedness can be easily solved by complexification of the path integral, which makes the path integral completely real-valued. The complexification, however, becomes increasingly complicated for anharmonic potentials. As a result, we will be inevitably forced to adapt the Lefschetz thimble/holomorphic flow approaches, which have their own complications \cite{Tranberg:1902.09147,Alexandru:1605.08040,Alexandru:1704.06404,leveragingML,complexNN}. A further problem with the DoS/LLR approach for continuous variables might be the loss of ergodicity for simulations with constrained values of the imaginary part of the action $S_I$, especially if one uses the Hybrid Monte-Carlo algorithm based on nearly continuous updates of field variables. Finally, the polynomial approximations to the density of states, which plays a crucial role in reducing statistical errors, might fail in the vicinity of phase transitions \cite{Lucini:1910.11026}.

    We should note that the DoS/LLR approach might also be useful for exploratory studies, for example, to guide the construction of analytic approximations. In fact, our work proceeded in exactly this way - we first measured the ``unsubtracted'' DoS function with LLR$_0$, and only after looking at the DoS plots in Fig.~\ref{fig:dos_llr0_vs_llr2} we realized how to subtract the leading and next-to-leading contributions specified in \cref{eq:no_flip,eq:single_flip}.

	\section*{Acknowledgements}
	This work was funded in part by the STFC Consolidated Grant ST/T000988/1.
	Numerical simulations were undertaken on Barkla, part of the High Performance Computing facilities at the University of Liverpool, UK as well as on
	the Cambridge Service for Data Driven Discovery (CSD3), part of which is operated by the University of Cambridge Research Computing on behalf of the STFC DiRAC HPC Facility (\url{www.dirac.ac.uk}). The DiRAC component of CSD3 was funded by BEIS capital funding via STFC capital grants ST/P002307/1 and ST/R002452/1 and STFC operations grant ST/R00689X/1. DiRAC is part of the National e-Infrastructure.
	We used the \texttt{R} library \texttt{hadron}~\cite{hadron} for data analysis.
	JO would like to thank the other participants of the Pollica Summer Workshop on Effective Field Theories 2022, in particular Michael Flynn, for many an insightful discussion.
	We also thank Kurt Langfeld, Paul Rakow and David Schaich for their valuable comments.
	
	\appendix
\section{Derivation of the classical Ising model in \texorpdfstring{$d+1$}{d+1} dimensions and further details}\label{sec:quantum_to_classical}
	The exponentials in equation~\eqref{eq:trotter_decomposition} can be calculated exactly in the canonical $z$-basis because the first term is diagonal and the second one can be decomposed into local blocks of size $2\times 2$ yielding
	\begin{align}
		\eto{\im\delta \mh_i\sigma_i^x} &=
		\matr{cc}{
			\cos{\delta \mh_i}&\im\sin{\delta \mh_i}\\
			\im\sin{\delta \mh_i}&\cos{\delta \mh_i}
		}\,.\label{eq:time_evolution_h}
	\end{align}

	Let us now compare these matrices with the local building blocks of the anisotropic classical Ising model with an arbitrary coupling matrix $J'$ within the first dimension and a nearest-neighbour coupling $h'$ within the second dimension defined by the action
	\begin{align}
		S &= S^1+S^2\,,\\
		S^1 &= -\sum_k\sum_{i,j}s_{i,k}J'_{ij}s_{j,k}\,,\\
		S^2 &= -\sum_i\sum_k h'_is_{i,k}s_{i,k+1}\,.
	\end{align}
	The physics of such a classical statistical system are governed by the Boltzmann weighted partition sum (over all possible spin configurations)
	\begin{align}
		Z &= \sum_{\{s\}} \eto{-S(s)}
	\end{align}
	inducing the transfer matrices
	\begin{align}
		T_1 &= \left(\eto{-S^1}\right)_{k,k+1}\\
		&= \mathrm{diag}\left(\left\{\sum_{i,j}s_{i}J'_{ij}s_{j}\, |\; s\in\{\pm1\}^L\right\}\right)
	\end{align}
	and
    \begin{widetext}
	\begin{eqnarray}
		(T_2)_i = \left(\eto{-(S^2)_i}\right)_{k,k+1}
		= \matr{cc}{\eto{h'_i} \eto{-h'_i}
        \\
        \eto{-h'_i} \eto{h'_i}}
		= \frac12 \matr{rr}{1&1\\1&-1}\matr{cc}{\eto{h'_i}+\eto{-h'_i}&0\\0&\eto{h'_i}-\eto{-h'_i}}\matr{rr}{1&1\\1&-1}\,.
    \label{eq:temporal_transfer_matrix}
	\end{eqnarray}
    \end{widetext}

	We immediately observe the similarity between $T_1$ and the interacting part of the time evolution operator. Therefore we set
	\begin{align}
		J' &\equiv \im\delta \mJ\,.
	\end{align}
	Note again that this $\mJ$ corresponds to the purely spatial coupling matrix. The coupling in temporal direction turns out to be more challenging as we have to exploit the similarity of $T_2$ with the  $2\times2$-matrix \eqref{eq:time_evolution_h}:
	\begin{align}
		\matr{cc}{\eto{ h'_i}&\eto{- h'_i}\\\eto{- h'_i}&\eto{ h'_i}} &\overset{!}{\propto}
		\matr{cc}{
			\cos{\delta \mh_i}&\im \sin{\delta \mh_i}\\
			\im \sin{\delta \mh_i}&\cos{\delta \mh_i}
		}\\
		\Rightarrow \eto{-2 h'_i} &= \im \tan \delta \mh_i
	\end{align}

	Together with the proportionality factor
	\begin{align}
		A &= \prod_i A_i^{N_t} \,,\\
		A_i &= \sqrt{\im\sin\delta \mh_i\cos\delta \mh_i}
	\end{align}
	this allows an exact identification between the trace of the quantum mechanical time evolution operator and the classical partition sum
	\begin{align}
		\tr U(t) &= A Z\,.\label{eq:classical_to_quantum}
	\end{align}

	We can define the purely real constants
	\begin{align}
		 J &\coloneqq \delta \mJ\,,\\
		 h_i &\coloneqq -\frac12 \log\tan \delta \mh_i\,,
	\end{align}
	so that the action reads
	\begin{eqnarray}
		S = -\im\sum_k\sum_{i,j}s_{i,k} J_{ij}s_{j,k}
		- \nonumber \\ - 
        \sum_i\sum_k\left( h_i - \frac\pi4\im\right) s_{i,k}s_{i,k+1}\,.
	\end{eqnarray}

	\subsection{Distribution of the imaginary part}
	It is instructive to investigate the distribution of $S_I$ before we go on. For this we define the auxiliary variable
	\begin{align}
		\theta_k &\coloneqq \sum_{i,j}s_{i,k} J_{ij}s_{j,k}
	\end{align}
	with the upper bound
	\begin{align}
		|\theta_k| &\le || J||_1\\
		&\le nL\, \delta\,||\mJ||_\infty\,,
	\end{align}
	where $n$ is the number of neighbours a single site can couple to ($n=2$ for pure nearest neighbour coupling in 1 dimension) and $||\cdot||_p$ denotes the $p$-norm. Since the considered systems is translationally invariant in the time direction, all $\theta_k$ are distributed identically, though not independently in general. The $\mathbb{Z}_2$ symmetry of spin reflections is unbroken, so the expectation value of $\theta_k$ is $\bar\theta_k = 0$.
	
	Suppose the $\theta_k$ were uncorrelated. Then according to the central limit theorem $S_I$ would follow a normal distribution
	\begin{align}
		S_I \sim \mathcal N (0,v)
	\end{align}
	with a variance
	\begin{align}
		v &\le N_t \left(nL\, \delta\,||\mJ||_\infty\right)^2\\
		&= \frac{1}{N_t}\left(nL\,t\,||\mJ||_\infty\right)^2
	\end{align}
	approaching zero in the continuum limit.
	
	In the opposite limit of maximal correlation on the other hand $S_I$ simplifies to
	\begin{align}
		S_I &= N_t \theta_k\\
		\Rightarrow |S_I| &\le N_t \, n L \delta \,||\mJ||_\infty\\
		&= nL\,t\,||\mJ||_\infty\,,
	\end{align}
	that is a constant in the time steps size. We do not take the case of anti-correlation into account since it is highly unphysical. Therefore we find that in any relevant case $S_I$ is bounded by the physical time extent and the variance is going to approach a constant
	\begin{align}
		\lim_{N_t\rightarrow\infty} v &\le t\,\xi\left(nL\,||\mJ||_\infty\right)^2
	\end{align}
	in the continuum limit, where $\xi$ is the correlation length in time direction. $\xi$ is proportional to the standard deviation of $S_I$.
	
	\subsection{Alternative boundary conditions}
	We remark here that the periodic boundary conditions (pbc) silently assumed above are the most usual but not the only possible choice. Physically they correspond to same-to-same scattering and are best suited for infinite time approximations. In contrast open boundary conditions (obc) correspond to all-to-all scattering and might be relevant in comparisons with matrix product state (MPS) or similar calculations where they are much easier to realise than pbc. Closed boundary conditions (cbc) allow for particular choices of initial and final states.
	
	In all the non-periodic cases there is one less time step than time slices, so the proportionality factor between quantum and classical partition sums reduces to $\prod_i A_i^{N_t-1}$. Furthermore the trace in equation~\eqref{eq:trace_partition_sum} has to be replaced by projections to corresponding vectors.
	
	For \textit{obc} one obtains
	\begin{align}
		Z_{R,i} &= \frac12 \left(1,1\right)(T_2)_i^{N_t} \matr{c}{1\\1}\\
		&= 2^{N_t-1}\cosh^{N_t-1} \tilde h_i'
	\end{align}
	and for \textit{cbc}
	\begin{align}
		Z_{R,i} &= \left(1,0\right)(T_2)_i^{N_t} \matr{c}{\cos\varphi\\\sin\varphi}\\
		\begin{split}
			&= 2^{N_t-1}\left(\left(\cos\varphi+\sin\varphi\right)\cosh^{N_t-1} \tilde h_i'\right.\\
			&\qquad\quad\left. + \left(\cos\varphi-\sin\varphi\right)\cosh^{N_t-1} \tilde h_i'\right)\,,
		\end{split}
	\end{align}
	where initial and final states are rotated against each other by the angle $\varphi$. We observe that, as expected, all the boundary conditions lead to the same result at infinite times $t\rightarrow\infty,\,N_t\rightarrow\infty$ and moreover the cbc case of $\varphi=0$ (same-to-same scattering) yields the same result as pbc up to an irrelevant difference of $N_t$ by one.
\section{Error scaling with bin size}\label{sec:error_with_bins}
	The discretisation of the DoS into bins as in equation~\eqref{eq:dos_to_bins} naturally leads to errors vanishing as the bin width goes to zero, but relevant for finite bin size. The observable we are ultimately interested in amounts to a Fourier transformation of the DoS, so we can judge the quality of the discretisation method by its ability to approximate the integral
	\begin{align}
		F_a(x_0,h) &\coloneqq \frac{1}{2h}\int\limits_{x_0-h}^{x_0+h}\md x \,\eto{f(x)+\im a x}\,.
	\end{align}

	The separate estimation of the DoS in bins of length $2h$ succeeded by a multiplication with the phase at the mid-point of the interval as employed in this work results in
	\begin{eqnarray}
		\frac{\eto{\im a x_0}}{2h}\int\limits_{x_0-h}^{x_0+h}\md x\, \eto{f(x)} 
        = \nonumber \\ = 
        F_a(x_0,h)\,\eto{-\frac 16\im a\left(2f'(x_0)+\im a\right)h^2 + \ordnung{h^4}}.
	\end{eqnarray}

	Alternative approximations include the classical mid-point formula
	\begin{eqnarray}
		\eto{f(x_0) + \im a x_0} 
        = \nonumber \\ =
        F_a(x_0,h)\,\eto{-\frac 16\left(f''(x_0)+\left(f'(x_0)+\im a\right)^2\right)h^2 + \ordnung{h^4}}
	\end{eqnarray}
	as well as a more intricate piecewise linear (trapezoidal) formula
	\begin{eqnarray}
		\frac{\eto{f(x_0)}}{2h}\int\limits_{x_0-h}^{x_0+h}\md x\, \eto{f'(x_0)x + \im a x} 
        = \nonumber\\ = 
        F_a(x_0,h)\,\eto{-\frac 16f''(x_0)h^2 + \ordnung{h^4}}
	\end{eqnarray}
	as used e.g.\ in Ref.~\cite{Langfeld:1509.08391}.
	
	Thus, all the different methods' multiplicative errors are of order $\ordnung{h^2}$. Empirically we find that the bin size can easily be chosen small enough to completely neglect this error since the uncertainties in the estimation of $f(x)$ itself dominate the total error.  	%
\section{LLR step size}\label{sec:llr_params}	
	First, we note that condition~\eqref{eq:beta_sum_infinite} is not sufficient in practice as the sum does not go up to infinity but only up to $\Lambda$. Therefore we require additionally
	\begin{align}
		\sum_{k=0}^\Lambda \beta_{N,M,\Lambda}(k) &\ge \sum_i\hat \alpha_i\,,\label{eq:beta_sum_to_lambda}
	\end{align}
	where we denote the exact logarithmic DoS with the hat $\hat\alpha_i$. The left hand side of equation~\eqref{eq:beta_sum_to_lambda} evaluates to
	\begin{eqnarray}
		\sum_{k=0}^\Lambda \beta_{N,M,\Lambda}(k) 
        = \nonumber \\ = 
        a_{N,M,\Lambda}\left(\ln\frac{\Lambda}{b_{N,M,\Lambda}}+\ordnung{b_{N,M,\Lambda}^{-1}}+\ordnung{\Lambda^{-1}}\right)
	\end{eqnarray}
	while the right hand side can only be approximated
	\begin{align}
		\sum_i\hat \alpha_i &\le \sum_i N\ln2\\
		&=MN\ln2\\
		&\approx \num{0.69} MN\,.
	\end{align}
	Though this is a conservative upper bound, it is not very far from realistic estimations. If we for instance assume a parabolic shape for the $\alpha_i$ with a maximum close to $N\ln2$, we obtain $\sum_i\hat \alpha_i \lesssim \frac23 \ln2\,MN\approx\num{0.46}MN$. Thus, up to a factor of order one, we find that
	\begin{align}
		a_{N,M,\Lambda} &= \ln2\,\frac{MN}{\ln\frac{\Lambda}{b_{N,M,\Lambda}}}
	\end{align}
	is a good choice.
	
	In order to get a value for the remaining offset $b_{N,M,\Lambda}$ we have to consider the small $k$ behaviour rather than the large $k$ limit as we did before because the offset is going to be irrelevant in the latter case. Consider the likely case that one of the first contributions $\beta_{N,M,\Lambda}(k\ll b_{N,M,\Lambda})$ is added to a state that has a much smaller exact $\hat\alpha_i$ relative to the other states. To compensate this `mistake' after some $k_0$ steps ($k_0$ can be large if the initial random walk remains in a distant region of the phase space for a while), we have to spend $\kappa$ steps adding this contribution to every other of the $M$ states. For this to be possible, the condition
	\begin{align}
		\frac{M}{b_{N,M,\Lambda}} &\le \sum_{k=k_0}^{k_0+\kappa}\frac{1}{b_{N,M,\Lambda}+k}\\
		&= \ln\frac{k_0+\kappa}{k_0} + \ordnung{\frac{b_{N,M,\Lambda}}{k_0}}
	\end{align}
	has to hold. This provides a lower bound for $\kappa$
	\begin{align}
		\kappa &\ge k_0\left(\eto{\frac{M}{b_{N,M,\Lambda}}}-1\right)
	\end{align}
	which has to be small for the algorithm to be efficient. Thus, it turns out that the offset $b_{N,M,\Lambda}\equiv b_M$ solely depends on the number of states and $b_M \gtrsim M$ for the runtime to not blow up exponentially. On the other hand we cannot choose $b_M$ arbitrarily large either since that would defeat the purpose of $\beta_{N,M,\Lambda}(k)$ decreasing quickly. Numerical test suggest that
	\begin{align}
		b_M &= 3M
	\end{align}
	yields a performance close to optimal.  	
	\bibliography{bibliography}

\begin{thebibliography}{47}%
\makeatletter
\providecommand \@ifxundefined [1]{%
 \@ifx{#1\undefined}
}%
\providecommand \@ifnum [1]{%
 \ifnum #1\expandafter \@firstoftwo
 \else \expandafter \@secondoftwo
 \fi
}%
\providecommand \@ifx [1]{%
 \ifx #1\expandafter \@firstoftwo
 \else \expandafter \@secondoftwo
 \fi
}%
\providecommand \natexlab [1]{#1}%
\providecommand \enquote  [1]{``#1''}%
\providecommand \bibnamefont  [1]{#1}%
\providecommand \bibfnamefont [1]{#1}%
\providecommand \citenamefont [1]{#1}%
\providecommand \href@noop [0]{\@secondoftwo}%
\providecommand \href [0]{\begingroup \@sanitize@url \@href}%
\providecommand \@href[1]{\@@startlink{#1}\@@href}%
\providecommand \@@href[1]{\endgroup#1\@@endlink}%
\providecommand \@sanitize@url [0]{\catcode `\\12\catcode `\$12\catcode
  `\&12\catcode `\#12\catcode `\^12\catcode `\_12\catcode `\%12\relax}%
\providecommand \@@startlink[1]{}%
\providecommand \@@endlink[0]{}%
\providecommand \url  [0]{\begingroup\@sanitize@url \@url }%
\providecommand \@url [1]{\endgroup\@href {#1}{\urlprefix }}%
\providecommand \urlprefix  [0]{URL }%
\providecommand \Eprint [0]{\href }%
\providecommand \doibase [0]{https://doi.org/}%
\providecommand \selectlanguage [0]{\@gobble}%
\providecommand \bibinfo  [0]{\@secondoftwo}%
\providecommand \bibfield  [0]{\@secondoftwo}%
\providecommand \translation [1]{[#1]}%
\providecommand \BibitemOpen [0]{}%
\providecommand \bibitemStop [0]{}%
\providecommand \bibitemNoStop [0]{.\EOS\space}%
\providecommand \EOS [0]{\spacefactor3000\relax}%
\providecommand \BibitemShut  [1]{\csname bibitem#1\endcsname}%
\let\auto@bib@innerbib\@empty
%</preamble>
\bibitem [{\citenamefont {Feynman}(1986)}]{Feynman:QuantMechComput}%
  \BibitemOpen
  \bibfield  {author} {\bibinfo {author} {\bibfnamefont {R.~P.}\ \bibnamefont
  {Feynman}},\ }\bibfield  {title} {\bibinfo {title} {Quantum mechanical
  computers},\ }\href {http://dx.doi.org/10.1007/BF01886518} {\bibfield
  {journal} {\bibinfo  {journal} {Found.~Phys.}\ }\textbf {\bibinfo {volume}
  {16}},\ \bibinfo {pages} {507 } (\bibinfo {year} {1986})}\BibitemShut
  {NoStop}%
\bibitem [{\citenamefont {Nagaj}(2010)}]{Nagaj:0908.4219}%
  \BibitemOpen
  \bibfield  {author} {\bibinfo {author} {\bibfnamefont {D.}~\bibnamefont
  {Nagaj}},\ }\bibfield  {title} {\bibinfo {title} {Fast universal quantum
  computation with railroad-switch local hamiltonians},\ }\href
  {http://dx.doi.org/10.1063/1.3384661} {\bibfield  {journal} {\bibinfo
  {journal} {J.~Math.~Phys.}\ }\textbf {\bibinfo {volume} {51}},\ \bibinfo
  {pages} {062201} (\bibinfo {year} {2010})},\ \Eprint
  {https://arxiv.org/abs/0908.4219} {0908.4219} \BibitemShut {NoStop}%
\bibitem [{\citenamefont {Kj\"all}\ \emph {et~al.}(2014)\citenamefont
  {Kj\"all}, \citenamefont {Bardarson},\ and\ \citenamefont
  {Pollmann}}]{PhysRevLett.113.107204}%
  \BibitemOpen
  \bibfield  {author} {\bibinfo {author} {\bibfnamefont {J.~A.}\ \bibnamefont
  {Kj\"all}}, \bibinfo {author} {\bibfnamefont {J.~H.}\ \bibnamefont
  {Bardarson}},\ and\ \bibinfo {author} {\bibfnamefont {F.}~\bibnamefont
  {Pollmann}},\ }\bibfield  {title} {\bibinfo {title} {{Many-Body Localization
  in a Disordered Quantum Ising Chain}},\ }\href
  {https://doi.org/10.1103/PhysRevLett.113.107204} {\bibfield  {journal}
  {\bibinfo  {journal} {Phys. Rev. Lett.}\ }\textbf {\bibinfo {volume} {113}},\
  \bibinfo {pages} {107204} (\bibinfo {year} {2014})},\ \Eprint
  {https://arxiv.org/abs/1403.1568} {1403.1568} \BibitemShut {NoStop}%
\bibitem [{\citenamefont {Serbyn}\ \emph {et~al.}(2015)\citenamefont {Serbyn},
  \citenamefont {Papi\'c},\ and\ \citenamefont {Abanin}}]{PhysRevX.5.041047}%
  \BibitemOpen
  \bibfield  {author} {\bibinfo {author} {\bibfnamefont {M.}~\bibnamefont
  {Serbyn}}, \bibinfo {author} {\bibfnamefont {Z.}~\bibnamefont {Papi\'c}},\
  and\ \bibinfo {author} {\bibfnamefont {D.~A.}\ \bibnamefont {Abanin}},\
  }\bibfield  {title} {\bibinfo {title} {{Criterion for Many-Body
  Localization-Delocalization Phase Transition}},\ }\href
  {https://doi.org/10.1103/PhysRevX.5.041047} {\bibfield  {journal} {\bibinfo
  {journal} {Phys. Rev. X}\ }\textbf {\bibinfo {volume} {5}},\ \bibinfo {pages}
  {041047} (\bibinfo {year} {2015})},\ \Eprint
  {https://arxiv.org/abs/1507.01635} {1507.01635} \BibitemShut {NoStop}%
\bibitem [{\citenamefont {Abanin}\ \emph {et~al.}(2021)\citenamefont {Abanin},
  \citenamefont {Bardarson}, \citenamefont {{De Tomasi}}, \citenamefont
  {Gopalakrishnan}, \citenamefont {Khemani}, \citenamefont {Parameswaran},
  \citenamefont {Pollmann}, \citenamefont {Potter}, \citenamefont {Serbyn},\
  and\ \citenamefont {Vasseur}}]{ABANIN2021168415}%
  \BibitemOpen
  \bibfield  {author} {\bibinfo {author} {\bibfnamefont {D.}~\bibnamefont
  {Abanin}}, \bibinfo {author} {\bibfnamefont {J.}~\bibnamefont {Bardarson}},
  \bibinfo {author} {\bibfnamefont {G.}~\bibnamefont {{De Tomasi}}}, \bibinfo
  {author} {\bibfnamefont {S.}~\bibnamefont {Gopalakrishnan}}, \bibinfo
  {author} {\bibfnamefont {V.}~\bibnamefont {Khemani}}, \bibinfo {author}
  {\bibfnamefont {S.}~\bibnamefont {Parameswaran}}, \bibinfo {author}
  {\bibfnamefont {F.}~\bibnamefont {Pollmann}}, \bibinfo {author}
  {\bibfnamefont {A.}~\bibnamefont {Potter}}, \bibinfo {author} {\bibfnamefont
  {M.}~\bibnamefont {Serbyn}},\ and\ \bibinfo {author} {\bibfnamefont
  {R.}~\bibnamefont {Vasseur}},\ }\bibfield  {title} {\bibinfo {title}
  {{Distinguishing localization from chaos: Challenges in finite-size
  systems}},\ }\href
  {https://doi.org/https://doi.org/10.1016/j.aop.2021.168415} {\bibfield
  {journal} {\bibinfo  {journal} {Ann.~Phys.}\ }\textbf {\bibinfo {volume}
  {427}},\ \bibinfo {pages} {168415} (\bibinfo {year} {2021})},\ \Eprint
  {https://arxiv.org/abs/1911.04501} {1911.04501} \BibitemShut {NoStop}%
\bibitem [{\citenamefont {Sels}\ and\ \citenamefont
  {Polkovnikov}(2021)}]{PhysRevE.104.054105}%
  \BibitemOpen
  \bibfield  {author} {\bibinfo {author} {\bibfnamefont {D.}~\bibnamefont
  {Sels}}\ and\ \bibinfo {author} {\bibfnamefont {A.}~\bibnamefont
  {Polkovnikov}},\ }\bibfield  {title} {\bibinfo {title} {Dynamical obstruction
  to localization in a disordered spin chain},\ }\href
  {https://doi.org/10.1103/PhysRevE.104.054105} {\bibfield  {journal} {\bibinfo
   {journal} {Phys.~Rev.~E}\ }\textbf {\bibinfo {volume} {104}},\ \bibinfo
  {pages} {054105} (\bibinfo {year} {2021})},\ \Eprint
  {https://arxiv.org/abs/2009.04501} {2009.04501} \BibitemShut {NoStop}%
\bibitem [{\citenamefont {Pietracaprina}\ \emph {et~al.}(2018)\citenamefont
  {Pietracaprina}, \citenamefont {Macé}, \citenamefont {Luitz},\ and\
  \citenamefont {Alet}}]{10.21468/SciPostPhys.5.5.045}%
  \BibitemOpen
  \bibfield  {author} {\bibinfo {author} {\bibfnamefont {F.}~\bibnamefont
  {Pietracaprina}}, \bibinfo {author} {\bibfnamefont {N.}~\bibnamefont
  {Macé}}, \bibinfo {author} {\bibfnamefont {D.~J.}\ \bibnamefont {Luitz}},\
  and\ \bibinfo {author} {\bibfnamefont {F.}~\bibnamefont {Alet}},\ }\bibfield
  {title} {\bibinfo {title} {{Shift-invert diagonalization of large many-body
  localizing spin chains}},\ }\href
  {https://doi.org/10.21468/SciPostPhys.5.5.045} {\bibfield  {journal}
  {\bibinfo  {journal} {SciPost~Phys.}\ }\textbf {\bibinfo {volume} {5}},\
  \bibinfo {pages} {45} (\bibinfo {year} {2018})},\ \Eprint
  {https://arxiv.org/abs/1803.05395} {1803.05395} \BibitemShut {NoStop}%
\bibitem [{\citenamefont {Sierant}\ \emph {et~al.}(2020)\citenamefont
  {Sierant}, \citenamefont {Lewenstein},\ and\ \citenamefont
  {Zakrzewski}}]{PhysRevLett.125.156601}%
  \BibitemOpen
  \bibfield  {author} {\bibinfo {author} {\bibfnamefont {P.}~\bibnamefont
  {Sierant}}, \bibinfo {author} {\bibfnamefont {M.}~\bibnamefont
  {Lewenstein}},\ and\ \bibinfo {author} {\bibfnamefont {J.}~\bibnamefont
  {Zakrzewski}},\ }\bibfield  {title} {\bibinfo {title} {Polynomially filtered
  exact diagonalization approach to many-body localization},\ }\href
  {https://doi.org/10.1103/PhysRevLett.125.156601} {\bibfield  {journal}
  {\bibinfo  {journal} {Phys.~Rev.~Lett.}\ }\textbf {\bibinfo {volume} {125}},\
  \bibinfo {pages} {156601} (\bibinfo {year} {2020})},\ \Eprint
  {https://arxiv.org/abs/2005.09534} {2005.09534} \BibitemShut {NoStop}%
\bibitem [{\citenamefont {Kiefer-Emmanouilidis}\ \emph
  {et~al.}(2021{\natexlab{a}})\citenamefont {Kiefer-Emmanouilidis},
  \citenamefont {Unanyan}, \citenamefont {Fleischhauer},\ and\ \citenamefont
  {Sirker}}]{PhysRevB.103.024203}%
  \BibitemOpen
  \bibfield  {author} {\bibinfo {author} {\bibfnamefont {M.}~\bibnamefont
  {Kiefer-Emmanouilidis}}, \bibinfo {author} {\bibfnamefont {R.}~\bibnamefont
  {Unanyan}}, \bibinfo {author} {\bibfnamefont {M.}~\bibnamefont
  {Fleischhauer}},\ and\ \bibinfo {author} {\bibfnamefont {J.}~\bibnamefont
  {Sirker}},\ }\bibfield  {title} {\bibinfo {title} {Slow delocalization of
  particles in many-body localized phases},\ }\href
  {https://link.aps.org/doi/10.1103/PhysRevB.103.024203} {\bibfield  {journal}
  {\bibinfo  {journal} {Phys.~Rev.~B}\ }\textbf {\bibinfo {volume} {103}},\
  \bibinfo {pages} {024203} (\bibinfo {year} {2021}{\natexlab{a}})},\ \Eprint
  {https://arxiv.org/abs/2010.00565} {2010.00565} \BibitemShut {NoStop}%
\bibitem [{\citenamefont {Kiefer-Emmanouilidis}\ \emph
  {et~al.}(2021{\natexlab{b}})\citenamefont {Kiefer-Emmanouilidis},
  \citenamefont {Unanyan}, \citenamefont {Fleischhauer},\ and\ \citenamefont
  {Sirker}}]{KIEFEREMMANOUILIDIS2021168481}%
  \BibitemOpen
  \bibfield  {author} {\bibinfo {author} {\bibfnamefont {M.}~\bibnamefont
  {Kiefer-Emmanouilidis}}, \bibinfo {author} {\bibfnamefont {R.}~\bibnamefont
  {Unanyan}}, \bibinfo {author} {\bibfnamefont {M.}~\bibnamefont
  {Fleischhauer}},\ and\ \bibinfo {author} {\bibfnamefont {J.}~\bibnamefont
  {Sirker}},\ }\bibfield  {title} {\bibinfo {title} {Unlimited growth of
  particle fluctuations in many-body localized phases},\ }\href
  {https://doi.org/10.1016/j.aop.2021.168481} {\bibfield  {journal} {\bibinfo
  {journal} {Ann.~Phys.}\ }\textbf {\bibinfo {volume} {435}},\ \bibinfo {pages}
  {168481} (\bibinfo {year} {2021}{\natexlab{b}})},\ \Eprint
  {https://arxiv.org/abs/2012.12436} {2012.12436} \BibitemShut {NoStop}%
\bibitem [{\citenamefont {Garron}\ and\ \citenamefont
  {Langfeld}(2017{\natexlab{a}})}]{Garron:1703.04649}%
  \BibitemOpen
  \bibfield  {author} {\bibinfo {author} {\bibfnamefont {N.}~\bibnamefont
  {Garron}}\ and\ \bibinfo {author} {\bibfnamefont {K.}~\bibnamefont
  {Langfeld}},\ }\bibfield  {title} {\bibinfo {title} {Controlling the sign
  problem in finite-density quantum field theory},\ }\href
  {https://doi.org/10.1140/epjc/s10052-017-5039-7} {\bibfield  {journal}
  {\bibinfo  {journal} {Eur.~Phys.~J.~C}\ }\textbf {\bibinfo {volume} {77}}
  (\bibinfo {year} {2017}{\natexlab{a}})},\ \Eprint
  {https://arxiv.org/abs/1703.04649} {1703.04649} \BibitemShut {NoStop}%
\bibitem [{\citenamefont {Garron}\ and\ \citenamefont
  {Langfeld}(2017{\natexlab{b}})}]{Garron:1611.01378}%
  \BibitemOpen
  \bibfield  {author} {\bibinfo {author} {\bibfnamefont {N.}~\bibnamefont
  {Garron}}\ and\ \bibinfo {author} {\bibfnamefont {K.}~\bibnamefont
  {Langfeld}},\ }\bibfield  {title} {\bibinfo {title} {Tackling the sign
  problem with a moment expansion and application to heavy dense {QCD}},\
  }\href {https://doi.org/10.22323/1.256.0084} {\bibfield  {journal} {\bibinfo
  {journal} {PoS}\ }\textbf {\bibinfo {volume} {LATTICE2016}},\ \bibinfo
  {pages} {084} (\bibinfo {year} {2017}{\natexlab{b}})},\ \Eprint
  {https://arxiv.org/abs/1611.01378} {1611.01378} \BibitemShut {NoStop}%
\bibitem [{\citenamefont {Bongiovanni}\ \emph {et~al.}(2016)\citenamefont
  {Bongiovanni}, \citenamefont {Langfeld}, \citenamefont {Lucini},
  \citenamefont {Pellegrini},\ and\ \citenamefont {Rago}}]{Rago:1601.02929}%
  \BibitemOpen
  \bibfield  {author} {\bibinfo {author} {\bibfnamefont {L.}~\bibnamefont
  {Bongiovanni}}, \bibinfo {author} {\bibfnamefont {K.}~\bibnamefont
  {Langfeld}}, \bibinfo {author} {\bibfnamefont {B.}~\bibnamefont {Lucini}},
  \bibinfo {author} {\bibfnamefont {R.}~\bibnamefont {Pellegrini}},\ and\
  \bibinfo {author} {\bibfnamefont {A.}~\bibnamefont {Rago}},\ }\bibfield
  {title} {\bibinfo {title} {The density of states approach at finite chemical
  potential: a numerical study of the {Bose} gas},\ }\href
  {https://doi.org/10.22323/1.251.0192} {\bibfield  {journal} {\bibinfo
  {journal} {PoS}\ }\textbf {\bibinfo {volume} {LATTICE2015}},\ \bibinfo
  {pages} {192} (\bibinfo {year} {2016})},\ \Eprint
  {https://arxiv.org/abs/1601.02929} {1601.02929} \BibitemShut {NoStop}%
\bibitem [{\citenamefont {Francesconi}\ \emph {et~al.}(2020)\citenamefont
  {Francesconi}, \citenamefont {Holzmann}, \citenamefont {Lucini},\ and\
  \citenamefont {Rago}}]{Lucini:1910.11026}%
  \BibitemOpen
  \bibfield  {author} {\bibinfo {author} {\bibfnamefont {O.}~\bibnamefont
  {Francesconi}}, \bibinfo {author} {\bibfnamefont {M.}~\bibnamefont
  {Holzmann}}, \bibinfo {author} {\bibfnamefont {B.}~\bibnamefont {Lucini}},\
  and\ \bibinfo {author} {\bibfnamefont {A.}~\bibnamefont {Rago}},\ }\bibfield
  {title} {\bibinfo {title} {Free energy of the self-interacting relativistic
  lattice {Bose} gas at finite density},\ }\href
  {http://dx.doi.org/10.1103/PhysRevD.101.014504} {\bibfield  {journal}
  {\bibinfo  {journal} {Phys.~Rev.~D}\ }\textbf {\bibinfo {volume} {101}},\
  \bibinfo {pages} {014504} (\bibinfo {year} {2020})},\ \Eprint
  {https://arxiv.org/abs/1910.11026} {1910.11026} \BibitemShut {NoStop}%
\bibitem [{\citenamefont {Wang}\ and\ \citenamefont
  {Landau}(2001)}]{PhysRevLett.86.2050}%
  \BibitemOpen
  \bibfield  {author} {\bibinfo {author} {\bibfnamefont {F.}~\bibnamefont
  {Wang}}\ and\ \bibinfo {author} {\bibfnamefont {D.~P.}\ \bibnamefont
  {Landau}},\ }\bibfield  {title} {\bibinfo {title} {{Efficient, Multiple-Range
  Random Walk Algorithm to Calculate the Density of States}},\ }\href
  {https://doi.org/10.1103/PhysRevLett.86.2050} {\bibfield  {journal} {\bibinfo
   {journal} {Phys.~Rev.~Lett.}\ }\textbf {\bibinfo {volume} {86}},\ \bibinfo
  {pages} {2050} (\bibinfo {year} {2001})},\ \Eprint
  {https://arxiv.org/abs/cond-mat/0011174} {cond-mat/0011174} \BibitemShut
  {NoStop}%
\bibitem [{\citenamefont {Belardinelli}\ and\ \citenamefont
  {Pereyra}(2007)}]{Wang_Landau_errors}%
  \BibitemOpen
  \bibfield  {author} {\bibinfo {author} {\bibfnamefont {R.~E.}\ \bibnamefont
  {Belardinelli}}\ and\ \bibinfo {author} {\bibfnamefont {V.~D.}\ \bibnamefont
  {Pereyra}},\ }\bibfield  {title} {\bibinfo {title} {{Wang-Landau algorithm: A
  theoretical analysis of the saturation of the error}},\ }\href
  {https://doi.org/10.1063/1.2803061} {\bibfield  {journal} {\bibinfo
  {journal} {J.~Chem.~Phys.}\ }\textbf {\bibinfo {volume} {127}},\ \bibinfo
  {pages} {184105} (\bibinfo {year} {2007})},\ \Eprint
  {https://arxiv.org/abs/cond-mat/0702414} {cond-mat/0702414} \BibitemShut
  {NoStop}%
\bibitem [{\citenamefont {Belardinelli}\ \emph {et~al.}(2008)\citenamefont
  {Belardinelli}, \citenamefont {Manzi},\ and\ \citenamefont
  {Pereyra}}]{PhysRevE.78.067701}%
  \BibitemOpen
  \bibfield  {author} {\bibinfo {author} {\bibfnamefont {R.~E.}\ \bibnamefont
  {Belardinelli}}, \bibinfo {author} {\bibfnamefont {S.}~\bibnamefont
  {Manzi}},\ and\ \bibinfo {author} {\bibfnamefont {V.~D.}\ \bibnamefont
  {Pereyra}},\ }\bibfield  {title} {\bibinfo {title} {{Analysis of the
  convergence of the $1/t$ and Wang-Landau algorithms in the calculation of
  multidimensional integrals}},\ }\href
  {https://doi.org/10.1103/PhysRevE.78.067701} {\bibfield  {journal} {\bibinfo
  {journal} {Phys.~Rev.~E}\ }\textbf {\bibinfo {volume} {78}},\ \bibinfo
  {pages} {067701} (\bibinfo {year} {2008})},\ \Eprint
  {https://arxiv.org/abs/0806.0268} {0806.0268} \BibitemShut {NoStop}%
\bibitem [{\citenamefont {Robbins}\ and\ \citenamefont
  {Monro}(1951)}]{RobbinsMonro}%
  \BibitemOpen
  \bibfield  {author} {\bibinfo {author} {\bibfnamefont {H.}~\bibnamefont
  {Robbins}}\ and\ \bibinfo {author} {\bibfnamefont {S.}~\bibnamefont
  {Monro}},\ }\bibfield  {title} {\bibinfo {title} {{A Stochastic Approximation
  Method}},\ }\href {https://doi.org/10.1214/aoms/1177729586} {\bibfield
  {journal} {\bibinfo  {journal} {Ann.~Math.~Stat.}\ }\textbf {\bibinfo
  {volume} {22}},\ \bibinfo {pages} {400 } (\bibinfo {year}
  {1951})}\BibitemShut {NoStop}%
\bibitem [{\citenamefont {Langfeld}\ \emph {et~al.}(2016)\citenamefont
  {Langfeld}, \citenamefont {Lucini}, \citenamefont {Pellegrini},\ and\
  \citenamefont {Rago}}]{Langfeld:1509.08391}%
  \BibitemOpen
  \bibfield  {author} {\bibinfo {author} {\bibfnamefont {K.}~\bibnamefont
  {Langfeld}}, \bibinfo {author} {\bibfnamefont {B.}~\bibnamefont {Lucini}},
  \bibinfo {author} {\bibfnamefont {R.}~\bibnamefont {Pellegrini}},\ and\
  \bibinfo {author} {\bibfnamefont {A.}~\bibnamefont {Rago}},\ }\bibfield
  {title} {\bibinfo {title} {An efficient algorithm for numerical computations
  of continuous densities of states},\ }\href
  {https://doi.org/10.1140/epjc/s10052-016-4142-5} {\bibfield  {journal}
  {\bibinfo  {journal} {Eur.~Phys.~J.~C}\ }\textbf {\bibinfo {volume} {76}},\
  \bibinfo {pages} {306} (\bibinfo {year} {2016})},\ \Eprint
  {https://arxiv.org/abs/1509.08391} {1509.08391} \BibitemShut {NoStop}%
\bibitem [{\citenamefont {K\"orner}\ \emph {et~al.}(2020)\citenamefont
  {K\"orner}, \citenamefont {Langfeld}, \citenamefont {Smith},\ and\
  \citenamefont {von Smekal}}]{PhysRevD.102.054502}%
  \BibitemOpen
  \bibfield  {author} {\bibinfo {author} {\bibfnamefont {M.}~\bibnamefont
  {K\"orner}}, \bibinfo {author} {\bibfnamefont {K.}~\bibnamefont {Langfeld}},
  \bibinfo {author} {\bibfnamefont {D.}~\bibnamefont {Smith}},\ and\ \bibinfo
  {author} {\bibfnamefont {L.}~\bibnamefont {von Smekal}},\ }\bibfield  {title}
  {\bibinfo {title} {{Density of states approach to the hexagonal Hubbard model
  at finite density}},\ }\href {https://doi.org/10.1103/PhysRevD.102.054502}
  {\bibfield  {journal} {\bibinfo  {journal} {Phys.~Rev.~D}\ }\textbf {\bibinfo
  {volume} {102}},\ \bibinfo {pages} {054502} (\bibinfo {year} {2020})},\
  \Eprint {https://arxiv.org/abs/2006.04607} {2006.04607} \BibitemShut
  {NoStop}%
\bibitem [{\citenamefont {Guagnelli}(2012)}]{Guagnelli:1209.4443}%
  \BibitemOpen
  \bibfield  {author} {\bibinfo {author} {\bibfnamefont {M.}~\bibnamefont
  {Guagnelli}},\ }\href@noop {} {\bibinfo {title} {Sampling the density of
  states}} (\bibinfo {year} {2012}),\ \Eprint {https://arxiv.org/abs/1209.4443}
  {1209.4443} \BibitemShut {NoStop}%
\bibitem [{\citenamefont {Cotler}\ \emph {et~al.}(2017)\citenamefont {Cotler},
  \citenamefont {{Hunter-Jones}}, \citenamefont {Liu},\ and\ \citenamefont
  {Yoshida}}]{Yoshida:1706.05400}%
  \BibitemOpen
  \bibfield  {author} {\bibinfo {author} {\bibfnamefont {J.}~\bibnamefont
  {Cotler}}, \bibinfo {author} {\bibfnamefont {N.}~\bibnamefont
  {{Hunter-Jones}}}, \bibinfo {author} {\bibfnamefont {J.}~\bibnamefont
  {Liu}},\ and\ \bibinfo {author} {\bibfnamefont {B.}~\bibnamefont {Yoshida}},\
  }\bibfield  {title} {\bibinfo {title} {Chaos, complexity, and random
  matrices},\ }\href {https://doi.org/10.1007/JHEP11(2017)048} {\bibfield
  {journal} {\bibinfo  {journal} {JHEP}\ }\textbf {\bibinfo {volume} {1711}},\
  \bibinfo {pages} {048}},\ \Eprint {https://arxiv.org/abs/1706.05400}
  {1706.05400} \BibitemShut {NoStop}%
\bibitem [{\citenamefont {Gharibyan}\ \emph {et~al.}(2018)\citenamefont
  {Gharibyan}, \citenamefont {Hanada}, \citenamefont {Shenker},\ and\
  \citenamefont {Tezuka}}]{Hanada:1803.08050}%
  \BibitemOpen
  \bibfield  {author} {\bibinfo {author} {\bibfnamefont {H.}~\bibnamefont
  {Gharibyan}}, \bibinfo {author} {\bibfnamefont {M.}~\bibnamefont {Hanada}},
  \bibinfo {author} {\bibfnamefont {S.~H.}\ \bibnamefont {Shenker}},\ and\
  \bibinfo {author} {\bibfnamefont {M.}~\bibnamefont {Tezuka}},\ }\bibfield
  {title} {\bibinfo {title} {Onset of random matrix behavior in scrambling
  systems},\ }\href {http://dx.doi.org/10.1007/JHEP07(2018)124} {\bibfield
  {journal} {\bibinfo  {journal} {JHEP}\ }\textbf {\bibinfo {volume} {1807}},\
  \bibinfo {pages} {124}},\ \Eprint {https://arxiv.org/abs/1803.08050}
  {1803.08050} \BibitemShut {NoStop}%
\bibitem [{\citenamefont {Liu}(2018)}]{Liu:1806.05316}%
  \BibitemOpen
  \bibfield  {author} {\bibinfo {author} {\bibfnamefont {J.}~\bibnamefont
  {Liu}},\ }\bibfield  {title} {\bibinfo {title} {Spectral form factors and
  late time quantum chaos},\ }\href
  {http://dx.doi.org/10.1103/PhysRevD.98.086026} {\bibfield  {journal}
  {\bibinfo  {journal} {Phys.~Rev.~D}\ }\textbf {\bibinfo {volume} {98}},\
  \bibinfo {pages} {086026} (\bibinfo {year} {2018})},\ \Eprint
  {https://arxiv.org/abs/1806.05316} {1806.05316} \BibitemShut {NoStop}%
\bibitem [{\citenamefont {Prakash}\ \emph {et~al.}(2021)\citenamefont
  {Prakash}, \citenamefont {Pixley},\ and\ \citenamefont
  {Kulkarni}}]{Prakash:2008.07547}%
  \BibitemOpen
  \bibfield  {author} {\bibinfo {author} {\bibfnamefont {A.}~\bibnamefont
  {Prakash}}, \bibinfo {author} {\bibfnamefont {J.~H.}\ \bibnamefont
  {Pixley}},\ and\ \bibinfo {author} {\bibfnamefont {M.}~\bibnamefont
  {Kulkarni}},\ }\bibfield  {title} {\bibinfo {title} {Universal spectral form
  factor for many-body localization},\ }\href
  {http://dx.doi.org/10.1103/PhysRevResearch.3.L012019} {\bibfield  {journal}
  {\bibinfo  {journal} {Phys.~Rev.~Research}\ }\textbf {\bibinfo {volume}
  {3}},\ \bibinfo {pages} {012019} (\bibinfo {year} {2021})},\ \Eprint
  {https://arxiv.org/abs/2008.07547} {2008.07547} \BibitemShut {NoStop}%
\bibitem [{\citenamefont {Vosk}\ \emph {et~al.}(2015)\citenamefont {Vosk},
  \citenamefont {Huse},\ and\ \citenamefont {Altman}}]{PhysRevX.5.031032}%
  \BibitemOpen
  \bibfield  {author} {\bibinfo {author} {\bibfnamefont {R.}~\bibnamefont
  {Vosk}}, \bibinfo {author} {\bibfnamefont {D.~A.}\ \bibnamefont {Huse}},\
  and\ \bibinfo {author} {\bibfnamefont {E.}~\bibnamefont {Altman}},\
  }\bibfield  {title} {\bibinfo {title} {{Theory of the Many-Body Localization
  Transition in One-Dimensional Systems}},\ }\href
  {https://doi.org/10.1103/PhysRevX.5.031032} {\bibfield  {journal} {\bibinfo
  {journal} {Phys. Rev. X}\ }\textbf {\bibinfo {volume} {5}},\ \bibinfo {pages}
  {031032} (\bibinfo {year} {2015})},\ \Eprint
  {https://arxiv.org/abs/1412.3117} {1412.3117} \BibitemShut {NoStop}%
\bibitem [{\citenamefont {Luitz}\ \emph {et~al.}(2015)\citenamefont {Luitz},
  \citenamefont {Laflorencie},\ and\ \citenamefont
  {Alet}}]{PhysRevB.91.081103}%
  \BibitemOpen
  \bibfield  {author} {\bibinfo {author} {\bibfnamefont {D.~J.}\ \bibnamefont
  {Luitz}}, \bibinfo {author} {\bibfnamefont {N.}~\bibnamefont {Laflorencie}},\
  and\ \bibinfo {author} {\bibfnamefont {F.}~\bibnamefont {Alet}},\ }\bibfield
  {title} {\bibinfo {title} {{Many-body localization edge in the random-field
  Heisenberg chain}},\ }\href {https://doi.org/10.1103/PhysRevB.91.081103}
  {\bibfield  {journal} {\bibinfo  {journal} {Phys. Rev. B}\ }\textbf {\bibinfo
  {volume} {91}},\ \bibinfo {pages} {081103} (\bibinfo {year} {2015})},\
  \Eprint {https://arxiv.org/abs/1411.0660} {1411.0660} \BibitemShut {NoStop}%
\bibitem [{\citenamefont {Abanin}\ \emph {et~al.}(2019)\citenamefont {Abanin},
  \citenamefont {Altman}, \citenamefont {Bloch},\ and\ \citenamefont
  {Serbyn}}]{RevModPhys.91.021001}%
  \BibitemOpen
  \bibfield  {author} {\bibinfo {author} {\bibfnamefont {D.~A.}\ \bibnamefont
  {Abanin}}, \bibinfo {author} {\bibfnamefont {E.}~\bibnamefont {Altman}},
  \bibinfo {author} {\bibfnamefont {I.}~\bibnamefont {Bloch}},\ and\ \bibinfo
  {author} {\bibfnamefont {M.}~\bibnamefont {Serbyn}},\ }\bibfield  {title}
  {\bibinfo {title} {{Colloquium: Many-body localization, thermalization, and
  entanglement}},\ }\href {https://dx.doi.org/10.1103/RevModPhys.91.021001}
  {\bibfield  {journal} {\bibinfo  {journal} {Rev. Mod. Phys.}\ }\textbf
  {\bibinfo {volume} {91}},\ \bibinfo {pages} {021001} (\bibinfo {year}
  {2019})},\ \Eprint {https://arxiv.org/abs/1804.11065} {1804.11065}
  \BibitemShut {NoStop}%
\bibitem [{\citenamefont {Gornyi}\ \emph {et~al.}(2005)\citenamefont {Gornyi},
  \citenamefont {Mirlin},\ and\ \citenamefont
  {Polyakov}}]{PhysRevLett.95.206603}%
  \BibitemOpen
  \bibfield  {author} {\bibinfo {author} {\bibfnamefont {I.~V.}\ \bibnamefont
  {Gornyi}}, \bibinfo {author} {\bibfnamefont {A.~D.}\ \bibnamefont {Mirlin}},\
  and\ \bibinfo {author} {\bibfnamefont {D.~G.}\ \bibnamefont {Polyakov}},\
  }\bibfield  {title} {\bibinfo {title} {Interacting electrons in disordered
  wires: {Anderson} localization and low-$t$ transport},\ }\href
  {https://doi.org/10.1103/PhysRevLett.95.206603} {\bibfield  {journal}
  {\bibinfo  {journal} {Phys. Rev. Lett.}\ }\textbf {\bibinfo {volume} {95}},\
  \bibinfo {pages} {206603} (\bibinfo {year} {2005})},\ \Eprint
  {https://arxiv.org/abs/cond-mat/0506411} {cond-mat/0506411} \BibitemShut
  {NoStop}%
\bibitem [{\citenamefont {Basko}\ \emph {et~al.}(2006)\citenamefont {Basko},
  \citenamefont {Aleiner},\ and\ \citenamefont {Altshuler}}]{BASKO20061126}%
  \BibitemOpen
  \bibfield  {author} {\bibinfo {author} {\bibfnamefont {D.}~\bibnamefont
  {Basko}}, \bibinfo {author} {\bibfnamefont {I.}~\bibnamefont {Aleiner}},\
  and\ \bibinfo {author} {\bibfnamefont {B.}~\bibnamefont {Altshuler}},\
  }\bibfield  {title} {\bibinfo {title} {Metal–insulator transition in a
  weakly interacting many-electron system with localized single-particle
  states},\ }\href {https://doi.org/10.1016/j.aop.2005.11.014} {\bibfield
  {journal} {\bibinfo  {journal} {Ann.~Phys.}\ }\textbf {\bibinfo {volume}
  {321}},\ \bibinfo {pages} {1126} (\bibinfo {year} {2006})},\ \Eprint
  {https://arxiv.org/abs/cond-mat/0506617} {cond-mat/0506617} \BibitemShut
  {NoStop}%
\bibitem [{\citenamefont {De~Roeck}\ and\ \citenamefont
  {Huveneers}(2017)}]{PhysRevB.95.155129}%
  \BibitemOpen
  \bibfield  {author} {\bibinfo {author} {\bibfnamefont {W.}~\bibnamefont
  {De~Roeck}}\ and\ \bibinfo {author} {\bibfnamefont {F.}~\bibnamefont
  {Huveneers}},\ }\bibfield  {title} {\bibinfo {title} {Stability and
  instability towards delocalization in many-body localization systems},\
  }\href {https://doi.org/10.1103/PhysRevB.95.155129} {\bibfield  {journal}
  {\bibinfo  {journal} {Phys. Rev. B}\ }\textbf {\bibinfo {volume} {95}},\
  \bibinfo {pages} {155129} (\bibinfo {year} {2017})},\ \Eprint
  {https://arxiv.org/abs/1608.01815} {1608.01815} \BibitemShut {NoStop}%
\bibitem [{\citenamefont {Sierant}\ and\ \citenamefont
  {Zakrzewski}(2022)}]{Sierant_2022}%
  \BibitemOpen
  \bibfield  {author} {\bibinfo {author} {\bibfnamefont {P.}~\bibnamefont
  {Sierant}}\ and\ \bibinfo {author} {\bibfnamefont {J.}~\bibnamefont
  {Zakrzewski}},\ }\bibfield  {title} {\bibinfo {title} {Challenges to
  observation of many-body localization},\ }\href
  {https://doi.org/10.1103/physrevb.105.224203} {\bibfield  {journal} {\bibinfo
   {journal} {Phys.~Rev.~B}\ }\textbf {\bibinfo {volume} {105}},\ \bibinfo
  {pages} {224203} (\bibinfo {year} {2022})},\ \Eprint
  {https://arxiv.org/abs/2109.13608} {2109.13608} \BibitemShut {NoStop}%
\bibitem [{\citenamefont {Atia}\ and\ \citenamefont
  {Aharonov}(2017)}]{Atia:1610.09619}%
  \BibitemOpen
  \bibfield  {author} {\bibinfo {author} {\bibfnamefont {Y.}~\bibnamefont
  {Atia}}\ and\ \bibinfo {author} {\bibfnamefont {D.}~\bibnamefont
  {Aharonov}},\ }\bibfield  {title} {\bibinfo {title} {Fast-forwarding of
  {Hamiltonians} and exponentially precise measurements},\ }\href
  {http://dx.doi.org/10.1038/s41467-017-01637-7} {\bibfield  {journal}
  {\bibinfo  {journal} {Nature Comm.}\ }\textbf {\bibinfo {volume} {8}},\
  \bibinfo {pages} {1572} (\bibinfo {year} {2017})},\ \Eprint
  {https://arxiv.org/abs/1610.09619} {1610.09619} \BibitemShut {NoStop}%
\bibitem [{\citenamefont {{R Core Team}}(2020)}]{r_language}%
  \BibitemOpen
  \bibfield  {author} {\bibinfo {author} {\bibnamefont {{R Core Team}}},\
  }\href {https://www.R-project.org/} {\emph {\bibinfo {title} {{R: A Language
  and Environment for Statistical Computing}}}},\ \bibinfo {organization} {R
  Foundation for Statistical Computing},\ \bibinfo {address} {Vienna, Austria}
  (\bibinfo {year} {2020})\BibitemShut {NoStop}%
\bibitem [{\citenamefont {Spiegelhalter}\ \emph {et~al.}(2002)\citenamefont
  {Spiegelhalter}, \citenamefont {Best}, \citenamefont {Carlin},\ and\
  \citenamefont {Van Der~Linde}}]{bayesian_measures}%
  \BibitemOpen
  \bibfield  {author} {\bibinfo {author} {\bibfnamefont {D.~J.}\ \bibnamefont
  {Spiegelhalter}}, \bibinfo {author} {\bibfnamefont {N.~G.}\ \bibnamefont
  {Best}}, \bibinfo {author} {\bibfnamefont {B.~P.}\ \bibnamefont {Carlin}},\
  and\ \bibinfo {author} {\bibfnamefont {A.}~\bibnamefont {Van Der~Linde}},\
  }\bibfield  {title} {\bibinfo {title} {Bayesian measures of model complexity
  and fit},\ }\href {https://doi.org/10.1111/1467-9868.00353} {\bibfield
  {journal} {\bibinfo  {journal} {J.~R.~Stat.~Soc.~B}\ }\textbf {\bibinfo
  {volume} {64}},\ \bibinfo {pages} {583} (\bibinfo {year} {2002})}\BibitemShut
  {NoStop}%
\bibitem [{\citenamefont {Nemirovskii}\ and\ \citenamefont
  {Yudin}(1983)}]{nemirovskii1983problem}%
  \BibitemOpen
  \bibfield  {author} {\bibinfo {author} {\bibfnamefont {A.}~\bibnamefont
  {Nemirovskii}}\ and\ \bibinfo {author} {\bibfnamefont {D.}~\bibnamefont
  {Yudin}},\ }\href@noop {} {\emph {\bibinfo {title} {Problem complexity and
  method efficiency in optimization}}}\ (\bibinfo  {publisher} {John Wiley \&
  Sons, New York},\ \bibinfo {year} {1983})\BibitemShut {NoStop}%
\bibitem [{\citenamefont {Ostmeyer}(2022)}]{j_ostmeyer_2022_7164902}%
  \BibitemOpen
  \bibfield  {author} {\bibinfo {author} {\bibfnamefont {J.}~\bibnamefont
  {Ostmeyer}},\ }\href {https://doi.org/10.5281/zenodo.7164902} {\bibinfo
  {title} {{j-ostmeyer/real-time-dos: Code and data along with
  arXiv:2209.13970v1}}} (\bibinfo {year} {2022})\BibitemShut {NoStop}%
\bibitem [{\citenamefont {Beard}\ and\ \citenamefont
  {Wiese}(1996)}]{PhysRevLett.77.5130}%
  \BibitemOpen
  \bibfield  {author} {\bibinfo {author} {\bibfnamefont {B.~B.}\ \bibnamefont
  {Beard}}\ and\ \bibinfo {author} {\bibfnamefont {U.-J.}\ \bibnamefont
  {Wiese}},\ }\bibfield  {title} {\bibinfo {title} {{Simulations of Discrete
  Quantum Systems in Continuous Euclidean Time}},\ }\href
  {https://doi.org/10.1103/PhysRevLett.77.5130} {\bibfield  {journal} {\bibinfo
   {journal} {Phys.~Rev.~Lett.}\ }\textbf {\bibinfo {volume} {77}},\ \bibinfo
  {pages} {5130} (\bibinfo {year} {1996})},\ \Eprint
  {https://arxiv.org/abs/cond-mat/9602164} {cond-mat/9602164} \BibitemShut
  {NoStop}%
\bibitem [{\citenamefont {Langfeld}\ and\ \citenamefont
  {Lucini}(2014)}]{Langfeld:1404.7187}%
  \BibitemOpen
  \bibfield  {author} {\bibinfo {author} {\bibfnamefont {K.}~\bibnamefont
  {Langfeld}}\ and\ \bibinfo {author} {\bibfnamefont {B.}~\bibnamefont
  {Lucini}},\ }\bibfield  {title} {\bibinfo {title} {The density of states
  approach to dense quantum systems},\ }\href
  {http://dx.doi.org/10.1103/PhysRevD.90.094502} {\bibfield  {journal}
  {\bibinfo  {journal} {Phys.~Rev.~D}\ }\textbf {\bibinfo {volume} {90}},\
  \bibinfo {pages} {094502} (\bibinfo {year} {2014})},\ \Eprint
  {https://arxiv.org/abs/1404.7187} {1404.7187} \BibitemShut {NoStop}%
\bibitem [{\citenamefont {Garron}\ and\ \citenamefont
  {Langfeld}(2016)}]{Langfeld:1605.02709}%
  \BibitemOpen
  \bibfield  {author} {\bibinfo {author} {\bibfnamefont {N.}~\bibnamefont
  {Garron}}\ and\ \bibinfo {author} {\bibfnamefont {K.}~\bibnamefont
  {Langfeld}},\ }\bibfield  {title} {\bibinfo {title} {Anatomy of the
  sign-problem in heavy-dense {QCD}},\ }\href
  {http://dx.doi.org/10.1140/epjc/s10052-016-4412-2} {\bibfield  {journal}
  {\bibinfo  {journal} {Eur.~Phys.~J.~C}\ }\textbf {\bibinfo {volume} {76}},\
  \bibinfo {pages} {569} (\bibinfo {year} {2016})},\ \Eprint
  {https://arxiv.org/abs/1605.02709} {1605.02709} \BibitemShut {NoStop}%
\bibitem [{\citenamefont {Schlüter}\ \emph {et~al.}(2021)\citenamefont
  {Schlüter}, \citenamefont {Gayk}, \citenamefont {Schmidt}, \citenamefont
  {Honecker},\ and\ \citenamefont {Schnack}}]{Schnack+2021}%
  \BibitemOpen
  \bibfield  {author} {\bibinfo {author} {\bibfnamefont {H.}~\bibnamefont
  {Schlüter}}, \bibinfo {author} {\bibfnamefont {F.}~\bibnamefont {Gayk}},
  \bibinfo {author} {\bibfnamefont {H.-J.}\ \bibnamefont {Schmidt}}, \bibinfo
  {author} {\bibfnamefont {A.}~\bibnamefont {Honecker}},\ and\ \bibinfo
  {author} {\bibfnamefont {J.}~\bibnamefont {Schnack}},\ }\bibfield  {title}
  {\bibinfo {title} {{Accuracy of the typicality approach using Chebyshev
  polynomials}},\ }\href {https://doi.org/doi:10.1515/zna-2021-0116} {\bibfield
   {journal} {\bibinfo  {journal} {Zeitschrift für Naturforschung A}\ }\textbf
  {\bibinfo {volume} {76}},\ \bibinfo {pages} {823} (\bibinfo {year} {2021})},\
  \Eprint {https://arxiv.org/abs/2104.13218} {2104.13218} \BibitemShut
  {NoStop}%
\bibitem [{\citenamefont {Mou}\ \emph {et~al.}(2019)\citenamefont {Mou},
  \citenamefont {Saffin}, \citenamefont {Tranberg},\ and\ \citenamefont
  {Woodward}}]{Tranberg:1902.09147}%
  \BibitemOpen
  \bibfield  {author} {\bibinfo {author} {\bibfnamefont {Z.-G.}\ \bibnamefont
  {Mou}}, \bibinfo {author} {\bibfnamefont {P.~M.}\ \bibnamefont {Saffin}},
  \bibinfo {author} {\bibfnamefont {A.}~\bibnamefont {Tranberg}},\ and\
  \bibinfo {author} {\bibfnamefont {S.}~\bibnamefont {Woodward}},\ }\bibfield
  {title} {\bibinfo {title} {Real-time quantum dynamics, path integrals and the
  method of thimbles},\ }\href {https://doi.org/10.1007/JHEP06(2019)094}
  {\bibfield  {journal} {\bibinfo  {journal} {JHEP}\ }\textbf {\bibinfo
  {volume} {1906}},\ \bibinfo {pages} {094}},\ \Eprint
  {https://arxiv.org/abs/1902.09147} {1902.09147} \BibitemShut {NoStop}%
\bibitem [{\citenamefont {Alexandru}\ \emph {et~al.}(2016)\citenamefont
  {Alexandru}, \citenamefont {Basar}, \citenamefont {Bedaque}, \citenamefont
  {Vartak},\ and\ \citenamefont {Warrington}}]{Alexandru:1605.08040}%
  \BibitemOpen
  \bibfield  {author} {\bibinfo {author} {\bibfnamefont {A.}~\bibnamefont
  {Alexandru}}, \bibinfo {author} {\bibfnamefont {G.}~\bibnamefont {Basar}},
  \bibinfo {author} {\bibfnamefont {P.~F.}\ \bibnamefont {Bedaque}}, \bibinfo
  {author} {\bibfnamefont {S.}~\bibnamefont {Vartak}},\ and\ \bibinfo {author}
  {\bibfnamefont {N.~C.}\ \bibnamefont {Warrington}},\ }\bibfield  {title}
  {\bibinfo {title} {{Monte Carlo} study of real time dynamics},\ }\href
  {https://doi.org/10.1103/PhysRevLett.117.081602} {\bibfield  {journal}
  {\bibinfo  {journal} {Phys.~Rev.~Lett.}\ }\textbf {\bibinfo {volume} {117}},\
  \bibinfo {pages} {081602} (\bibinfo {year} {2016})},\ \Eprint
  {https://arxiv.org/abs/1605.08040} {1605.08040} \BibitemShut {NoStop}%
\bibitem [{\citenamefont {Alexandru}\ \emph {et~al.}(2017)\citenamefont
  {Alexandru}, \citenamefont {Basar}, \citenamefont {Bedaque},\ and\
  \citenamefont {Ridgway}}]{Alexandru:1704.06404}%
  \BibitemOpen
  \bibfield  {author} {\bibinfo {author} {\bibfnamefont {A.}~\bibnamefont
  {Alexandru}}, \bibinfo {author} {\bibfnamefont {G.}~\bibnamefont {Basar}},
  \bibinfo {author} {\bibfnamefont {P.~F.}\ \bibnamefont {Bedaque}},\ and\
  \bibinfo {author} {\bibfnamefont {G.~W.}\ \bibnamefont {Ridgway}},\
  }\bibfield  {title} {\bibinfo {title} {{Schwinger-Keldysh} on the lattice: a
  faster algorithm and its application to field theory},\ }\href
  {http://dx.doi.org/10.1103/PhysRevD.95.114501} {\bibfield  {journal}
  {\bibinfo  {journal} {Phys.~Rev.~D}\ }\textbf {\bibinfo {volume} {95}},\
  \bibinfo {pages} {114501} (\bibinfo {year} {2017})},\ \Eprint
  {https://arxiv.org/abs/1704.06404} {1704.06404} \BibitemShut {NoStop}%
\bibitem [{\citenamefont {Wynen}\ \emph {et~al.}(2021)\citenamefont {Wynen},
  \citenamefont {Berkowitz}, \citenamefont {Krieg}, \citenamefont {Luu},\ and\
  \citenamefont {Ostmeyer}}]{leveragingML}%
  \BibitemOpen
  \bibfield  {author} {\bibinfo {author} {\bibfnamefont {J.-L.}\ \bibnamefont
  {Wynen}}, \bibinfo {author} {\bibfnamefont {E.}~\bibnamefont {Berkowitz}},
  \bibinfo {author} {\bibfnamefont {S.}~\bibnamefont {Krieg}}, \bibinfo
  {author} {\bibfnamefont {T.}~\bibnamefont {Luu}},\ and\ \bibinfo {author}
  {\bibfnamefont {J.}~\bibnamefont {Ostmeyer}},\ }\bibfield  {title} {\bibinfo
  {title} {{Machine learning to alleviate Hubbard-model sign problems}},\
  }\href {https://doi.org/10.1103/PhysRevB.103.125153} {\bibfield  {journal}
  {\bibinfo  {journal} {Phys.~Rev.~B}\ }\textbf {\bibinfo {volume} {103}},\
  \bibinfo {pages} {125153} (\bibinfo {year} {2021})},\ \Eprint
  {https://arxiv.org/abs/2006.11221} {2006.11221} \BibitemShut {NoStop}%
\bibitem [{\citenamefont {Rodekamp}\ \emph {et~al.}(2022)\citenamefont
  {Rodekamp}, \citenamefont {Berkowitz}, \citenamefont {Gäntgen},
  \citenamefont {Krieg}, \citenamefont {Luu},\ and\ \citenamefont
  {Ostmeyer}}]{complexNN}%
  \BibitemOpen
  \bibfield  {author} {\bibinfo {author} {\bibfnamefont {M.}~\bibnamefont
  {Rodekamp}}, \bibinfo {author} {\bibfnamefont {E.}~\bibnamefont {Berkowitz}},
  \bibinfo {author} {\bibfnamefont {C.}~\bibnamefont {Gäntgen}}, \bibinfo
  {author} {\bibfnamefont {S.}~\bibnamefont {Krieg}}, \bibinfo {author}
  {\bibfnamefont {T.}~\bibnamefont {Luu}},\ and\ \bibinfo {author}
  {\bibfnamefont {J.}~\bibnamefont {Ostmeyer}},\ }\bibfield  {title} {\bibinfo
  {title} {{Mitigating the Hubbard Sign Problem with Complex-Valued Neural
  Networks}},\ }\href {https://doi.org/10.1103/PhysRevB.106.125139} {\bibfield
  {journal} {\bibinfo  {journal} {Phys. Rev. B}\ }\textbf {\bibinfo {volume}
  {106}},\ \bibinfo {pages} {125139} (\bibinfo {year} {2022})},\ \Eprint
  {https://arxiv.org/abs/2203.00390} {2203.00390} \BibitemShut {NoStop}%
\bibitem [{\citenamefont {Kostrzewa}\ \emph {et~al.}(2020)\citenamefont
  {Kostrzewa}, \citenamefont {Ostmeyer}, \citenamefont {Ueding},\ and\
  \citenamefont {Urbach}}]{hadron}%
  \BibitemOpen
  \bibfield  {author} {\bibinfo {author} {\bibfnamefont {B.}~\bibnamefont
  {Kostrzewa}}, \bibinfo {author} {\bibfnamefont {J.}~\bibnamefont {Ostmeyer}},
  \bibinfo {author} {\bibfnamefont {M.}~\bibnamefont {Ueding}},\ and\ \bibinfo
  {author} {\bibfnamefont {C.}~\bibnamefont {Urbach}},\ }\href
  {https://github.com/HISKP-LQCD/hadron} {\emph {\bibinfo {title} {{hadron:
  Analysis Framework for Monte Carlo Simulation Data in Physics}}}} (\bibinfo
  {year} {2020}),\ \bibinfo {note} {r package version 3.1.2}\BibitemShut
  {NoStop}%
\end{thebibliography}%
\end{document}